%% file: main.tex
\documentclass[10pt,letterpaper]{article}
\usepackage[textwidth=6.1in,top=1in,bottom=1in]{geometry}
\usepackage{times}
\usepackage[authoryear,round]{natbib}
\usepackage[T1]{fontenc}
\usepackage[utf8]{inputenc}
\usepackage{microtype}
\usepackage{amsmath,amssymb,amsthm,mathtools}
\usepackage{graphicx,booktabs,array,multirow}
\usepackage{xcolor}
\usepackage{float,tabularx}
\floatstyle{ruled}\newfloat{algorithm}{htbp}{loa}\floatname{algorithm}{Algorithm}
\usepackage{hyperref}
\usepackage{url}
\usepackage{placeins}
\usepackage{fvextra}
\hypersetup{hidelinks,pdfauthor={Wenzhang Du},pdftitle={When Is Program Structure Required? Certificate-Grounded Calibration of Program-Induction Difficulty}}
\newtheorem{lemma}{Lemma}
\newtheorem{proposition}{Proposition}
\newtheorem{corollary}{Corollary}
\newcommand{\Lmin}{L_{\min}}
\newcommand{\rep}{\operatorname{replace}}
\newcommand{\Sone}{\mathcal{S}_1}
\newcommand{\ind}{\mathbf{1}}
\newcommand{\pp}{\,\mathrm{pp}}

\newcommand{\Com}{\mathrm{com}}

\title{When Is Program Structure Required?\\[0.35em]{\large Certificate-Grounded Calibration of Program-Induction Difficulty}}
\author{Wenzhang Du\\[0.35em]\normalsize Independent Researcher, China}
\date{}
\begin{document}
\maketitle

\begin{abstract}
A program-induction benchmark is informative when its examples require the structure it aims to test. Yet a two-rule generator can produce demonstrations that a single rule also explains. We introduce certificate-grounded calibration, which constructs matched tasks with minimum solution lengths of one and two in a bounded string-replacement language. Complete one-rule enumeration and an executable two-rule witness certify the change before model inference. We evaluate 120 reference families with three to six shared examples, using 1,440 new model trajectories and 480 saved trajectories. Across Qwen3-8B and Olmo-3-7B-Think, displayed-task completion is 23.53--45.83 percentage points lower on certified separation tasks, while success on a wholly unshown witness pair is 8.40--13.45 points higher. These witness gains accompany lower agreement on uniformly sampled strings and 1.93--5.78\% agreement on inputs that activate both reference rules. Recertification also reveals that a single unchanged example can exclude a shortcut and determine whether two rules are required. Together, these results show how the evidence supplied by examples shapes both the required solution and the programs models produce. The construction provides an executable testbed for comparing program search, final-answer delivery, and behavior on new inputs under verified task requirements.
\end{abstract}

\section{Introduction}\label{sec:intro}
A user can specify a text transformation through a few input-output examples. A program-induction system must then infer operations that fit the demonstrations and apply the intended transformation to new inputs. Programming-by-example systems use this ability for tasks such as spreadsheet transformations \citep{gulwani2011,gulwani2017}. Evaluating it requires demonstrations that actually demand the program structure a benchmark aims to test.

\begin{figure}[!tp]
\centering\includegraphics[width=\linewidth]{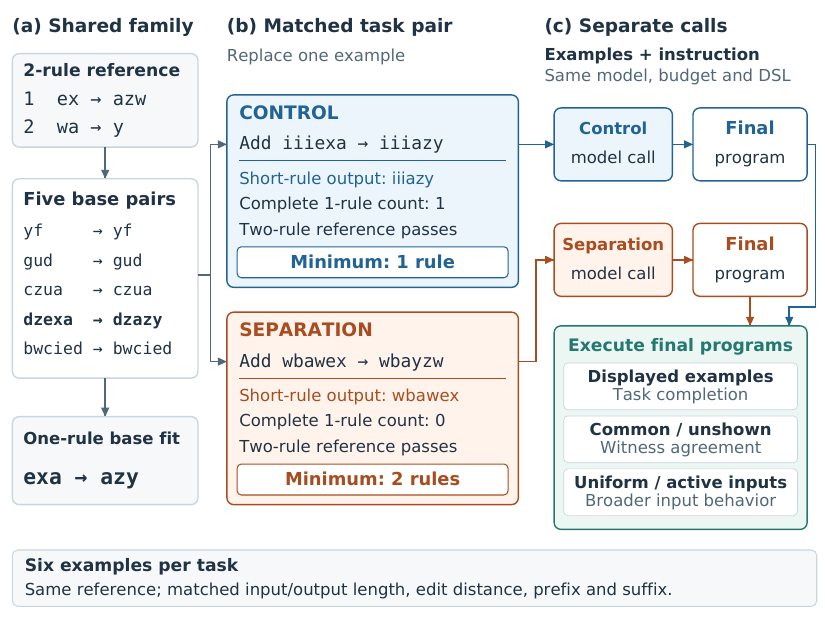}
\caption{\textbf{Certifying the program structure required by an example set.} (a) A two-rule reference generates five shared examples, all of which also admit the single replacement \texttt{exa}$\to$\texttt{azy}. (b) Two matched additions produce tasks with minimum lengths of one and two rules. Complete enumeration in the stated replacement language verifies the one-rule solution sets, and the reference satisfies both tasks. (c) Each model call receives the displayed examples and a synthesis instruction. The returned program is executed on its displayed task, the common-nine target, the wholly unshown witness pair, and separately sampled uniform and rule-active inputs.}
\label{fig:construction}
\end{figure}

Consider a program that first replaces \texttt{ex} with \texttt{azw} and then replaces \texttt{wa} with \texttt{y}. It maps \texttt{dzexa} to \texttt{dzazy}. Together with four unchanged examples, this mapping admits exactly one one-rule solution in our bounded replacement language: \texttt{exa}$\to$\texttt{azy}. Adding \texttt{iiiexa}$\to$\texttt{iiiazy} preserves that solution. Replacing the added example with \texttt{wbawex}$\to$\texttt{wbayzw} excludes it while retaining the two-rule reference. The two additions have equal input and output lengths and equal edit features. Figure~\ref{fig:construction} follows this change from the examples to the required solution length.

This ambiguity matters when benchmark scores are used to assess multi-step program induction. A model can complete examples generated by two rules using a single rule whenever the demonstrations allow it. In a saved archive of 1,008 replacement tasks, complete one-rule enumeration identifies 23 tasks admitting a shorter one-rule solution. A direct check of the minimum consistent length makes the structure demanded by each task explicit. It also creates an empirical question: how do model outputs change when the demonstrations require an additional rule, and how do those outputs behave on new inputs?

We answer this question with \emph{certificate-grounded calibration}. For a shared base of examples, we enumerate the complete set of legal one-rule solutions. A control addition preserves a short solution; a separation addition excludes the entire one-rule class. A verified two-rule reference satisfies both specifications. These checks certify minimum lengths of exactly one and two while preserving the reference program, the number of demonstrations, and five surface features of the changed example. The certification happens before model inference.

The evaluation follows each returned program from its displayed task to several sets of new inputs. On 120 new reference families, Qwen3-8B and Olmo-3-7B-Think complete substantially fewer separation tasks, yet more often satisfy a common witness pair that neither prompt contains. Extending the same families from three to six common examples preserves both directions. On separately sampled inputs, performance depends strongly on the evaluation distribution: uniform-input agreement decreases throughout the panel, and agreement on rule-active inputs remains low. A symbolic search baseline and saved-trajectory analyses help locate these observations in the generated programs.

Our contributions are threefold. First, we certify which program structure the displayed examples require, rather than inferring it from the generating reference. Second, we show how the same certified change accompanies different outcomes on task completion, distinguishing witnesses, and broader inputs across model and example-count settings. Third, we use the certificates to reassess saved tasks and to identify an unchanged example that rules out a shorter explanation. The companion artifact provides executable certificates, saved Qwen/Olmo outputs, evaluation targets, and CPU replay materials.

\section{Related Work}
\paragraph{Synthesis from examples.}
FlashFill resolves ambiguities in string transformations through examples \citep{gulwani2011}. Counterexample-guided synthesis in SKETCH \citep{solar2006,solar2008}, syntax-guided synthesis \citep{alur2013}, and query learning \citep{angluin1987} use evidence to restrict candidate programs. SemGuS supports user-defined semantics and synthesis or unrealizability reasoning through constrained Horn clauses \citep{kim2021semgus}. Our construction uses complete exclusion of a bounded short-program class to produce benchmark items with verified solution requirements.

\paragraph{Neural program induction.}
DeepCoder guides symbolic search with learned program properties \citep{balog2017}; RobustFill learns string programs from examples \citep{devlin2017}. ExeDec decomposes synthesis through execution subgoals \citep{shi2024exedec}, and Latent Programmer organizes search with discrete program codes \citep{hong2021latent}. ReaComp compiles model reasoning into symbolic solvers \citep{reacomp}. These approaches study how to find programs. Certifying the structure required by their input specifications provides a complementary way to organize and interpret evaluation.

\paragraph{Executable benchmarks and measurement.}
PBEBench controls generated replacement cascades and studies shorter predictions and bypassed relations \citep{pbebench}. Our checker quantifies over every legal one-rule solution to the displayed examples, enabling a certificate before predictions are collected. ReplaySCM uses executable mechanisms and intervention replay \citep{replayscm}; F-ICL supplies an exact bounded-program reference and audits runtime and template effects \citep{ficl}. Work on shortcut learning \citep{geirhos2020}, annotation artifacts \citep{gururangan2018}, HANS \citep{mccoy2019}, CheckList \citep{ribeiro2020}, and contrast sets \citep{gardner2020} motivates evaluations that expose the behavior underlying aggregate scores. Measurement modeling makes the link from scores to intended constructs explicit \citep{jacobs2021,bowman2021}. Our minimum-length certificates provide one executable link for program induction. Chain-of-thought and broad capability evaluations motivate the study of multi-step behavior \citep{wei2022,srivastava2023}; prefix-level evaluations study continued reasoning \citep{thinkingpast}. We use saved prefixes to measure when correct candidate programs and final answers become available.

\section{Certifying Required Program Length}\label{sec:certificate}
\subsection{Language and complete one-rule enumeration}
Let $\Sigma$ be a finite alphabet. A rule $\rep(a,b)$ replaces all nonoverlapping occurrences of nonempty $a$ by $b$ in one pass, using Python string-replacement semantics. A program applies its rules in order. The experimental language $\mathcal P$ contains one to five rules over $\Sigma=\{a,\ldots,k,u,v,w,x,y,z\}$, with $1\leq |a|\leq3$ and $0\leq |b|\leq3$.

For an example set $E=\{(x_i,y_i)\}_{i=1}^n$, write $p\models E$ when $p(x_i)=y_i$ for every example, and define
\begin{equation}
\Lmin(E)=\min_{p\in\mathcal P:\,p\models E}|p|.
\end{equation}
Our constructed tasks contain a nonidentity mapping and a valid two-rule reference, so this minimum exists. To compute the one-rule solution set $\Sone(E)$, let $\operatorname{Sub}_{1:k}(X)$ collect substrings of lengths one through $k$ from strings in $X$. We execute all candidates in
\begin{equation}
A(E)\times B(E),\qquad
A(E)=\operatorname{Sub}_{1:k_a}(\{x_i\}),\quad
B(E)=\operatorname{Sub}_{1:k_b}(\{y_i\})\cup\{\epsilon\},
\end{equation}
and retain the candidates that satisfy every example. Experiments use $k_a=k_b=3$.

\begin{lemma}[One-rule completeness]\label{lem:complete}
For a specification containing a nonidentity mapping, this procedure enumerates every legal one-rule solution under fixed finite bounds $k_a,k_b$.
\end{lemma}
\begin{proof}
A solution $\rep(a,b)$ acts on a nonidentity example, so $a$ occurs in an input and belongs to $A(E)$. If $b$ is empty it is explicitly enumerated. Otherwise, an inserted copy of $b$ remains a contiguous substring of the single-pass output, placing $b$ in $B(E)$. The candidate is therefore tested and retained. Each retained candidate is legal and directly checked on every example.
\end{proof}

\begin{proposition}[Certified one-to-two switch]\label{prop:switch}
If nonidentity specifications $E_C,E_S$ share a valid two-rule reference, $\Sone(E_C)\ne\varnothing$, and $\Sone(E_S)=\varnothing$, then $\Lmin(E_C)=1$ and $\Lmin(E_S)=2$.
\end{proposition}
\begin{proof}
A retained one-rule program attains the control minimum. Lemma~\ref{lem:complete} excludes every one-rule separation solution, and the shared reference attains length two.
\end{proof}

\subsection{Matched specifications and witnesses}
A family $f$ contains a two-rule reference $p_f^*$ and five base examples $B_f$. References come from the feeding-only class of the PBEBench generator. We retain nonidentity bases with $\Sone(B_f)\ne\varnothing$. Additional examples are produced by executing $p_f^*$. A control witness preserves a base one-rule solution; a separation witness rejects all of them.

Each accepted family contains two matched witness pairs $(c_1,s_1),(c_2,s_2)$. The four witnesses are nonidentity mappings with distinct inputs absent from the base. Within each pair, we exactly match
\begin{equation}
s(x,y)=(|x|,|y|,d_{\rm edit}(x,y),\ell_{\rm prefix},\ell_{\rm suffix}),
\end{equation}
where the common suffix excludes overlap with the common prefix. A predetermined choice $j\in\{1,2\}$ gives
\begin{equation}
E_{fC}=B_f\cup\{c_j\},\qquad E_{fS}=B_f\cup\{s_j\}.\label{eq:conditions}
\end{equation}
Both tasks contain six examples. We independently re-enumerate their one-rule solutions and execute the reference on all examples. Figure~\ref{fig:construction} gives a complete base and one matched pair; Appendix~\ref{app:construction} supplies the other pair and construction procedure.

The resulting task pair preserves the reference, common base, example count, and matched edit features. Its changed example carries the evidence that excludes the short solution class. This gives a useful prediction about solvers that prefer short consistent programs.

\begin{corollary}[Short solutions and separating witnesses]\label{cor:short}
Every one-rule solution to $E_{fC}$ fails each separation witness $s_1,s_2$.
\end{corollary}
\begin{proof}
Such a program belongs to $\Sone(B_f)$, and each separation witness rejects every member of that set.
\end{proof}
A shortest-first solver therefore completes control tasks with programs that fail separating witnesses. A two-rule reference satisfies every witness. Evaluating returned programs on these examples reveals how a solver resolves the ambiguity left by the displayed task.

\subsection{A certificate interface for other task classes}\label{sec:generalization}
The construction can be specified through four objects: an executable program class $\mathcal P$, a structural cost $c(p)$, a complete checker for solutions below a chosen cost, and a verified program attaining that cost. Given a task $E$, exclusion of every $p$ with $c(p)<k$ and a valid cost-$k$ witness certifies its minimum cost. Matched tasks can then be built around the same reference and evaluated under a common inference protocol.

For enumeration, assume that every cost-bounded class $\{p\in\mathcal P:c(p)\leq k\}$ is finite and effectively enumerable, and that execution terminates on the task inputs. Enumerating the lower-cost programs and executing them on the examples then yields a complete check. Retaining the consistent programs gives the short-solution set. The reference supplies candidate additions; an addition is accepted when it excludes every surviving short program. The certificate and solver evaluation use the same grammar, interpreter, and structural cost.

For a two-to-three transition, the control requires a valid two-rule program and complete exclusion of one-rule solutions; the separation requires a valid three-rule program and complete exclusion of all programs of length at most two. A checker for two-rule programs must cover intermediate strings created by the first replacement and consumed by the second. Finite alphabets and bounded rule strings define an exhaustive search domain. Applying this interface to another synthesis benchmark requires an appropriate complete lower-cost decision procedure and a feasible construction of the paired specifications.

\section{Study Design}\label{sec:design}
\subsection{Families and example-count extension}
Two discovery studies used 60 families and 480 outputs each. They motivated an independent validation on 120 new reference families, accepted from 15,004 proposals under fixed eligibility rules. Reference-program identity checks exclude the earlier references, proposals, predictions, and constructive alternatives. The base has one, two, or three one-rule solutions in 91, 13, and 16 families. Witness assignment and first-executed condition are balanced 60/60, and paired prompts use the same example permutation.

The extension evaluates the same families with $b\in\{3,4,5,6\}$ common examples and $b+1$ displayed examples. Reduced bases retain a fixed nonidentity anchor and use a fixed input-hash order to form nested subsets. The extra base-six example satisfies both the one-rule control program and the family's two-rule reference. It is inserted immediately before the original changed example, preserving the original examples' relative order. Each variant is recertified before inference.

At base counts three and four, 119 families retain a one-to-two certificate and one family admits a one-rule separation solution. At counts five and six, all 120 retain the one-to-two certificate. Table~\ref{tab:basecounts} reports the certified comparisons; Appendices~\ref{app:r1outcomes} and~\ref{app:r1oneone} give the full panel and the one-to-one case. The number of base one-rule solutions is one, two, or three in 90/13/17 families at counts three and four, and 91/13/16 at counts five and six. Appendix~\ref{app:r1variants} summarizes base composition and solution counts; the supplement contains the enumerated solution sets.

\subsection{Models and returned programs}
The original validation evaluates Qwen3-8B \citep{qwen2025}, Olmo-3-7B-Think \citep{olmo2025}, and Phi-4-mini-reasoning \citep{phi2025}. The extension evaluates Qwen and Olmo, reusing their 480 base-five trajectories and adding 1,440 trajectories at counts three, four, and six. Each model--family--condition--count cell contains one sampled trajectory. Model revisions, prompts, generation settings, and final-program selectors are fixed before the corresponding inference run.

Models use BF16 weights, temperature 0.6, top-$p$ 0.95, and a maximum of 8,192 new tokens. Qwen uses top-$k$ 20 and batch size eight; Olmo and Phi disable top-$k$ and use batch size four. Each prompt supplies the displayed examples and an instruction to return a legal cascade as a Python list, with reasoning permitted first. Appendix~\ref{app:implementation} gives the verbatim instruction and pinned configurations.

The canonical selector maps a fixed set of rule notations to legal ordered replacements. For Qwen and Olmo it selects the last complete Python fence after the final closed thinking delimiter, falling back to the last complete list when that fence is absent. Phi uses the same grammar with its fixed endpoint adapter. Selection occurs before program execution. We also record whether any legal candidate expressed in a trajectory satisfies the task, and compare canonical execution with the original native extractor. These measurements track final-answer correctness and correct-candidate availability.

\subsection{Evaluation inputs and uncertainty}
We evaluate the selected program $\hat p_{fcb}$ on its displayed specification and on shared reference targets. The original common-nine target is
\begin{equation}
E_f^{\Com}=B_f\cup\{c_1,s_1,c_2,s_2\}.
\end{equation}
The wholly unshown witness pair is
\begin{equation}
H_{2,f}=\{c_{3-j},s_{3-j}\}.\label{eq:h2}
\end{equation}
Both examples in $H_{2,f}$ are absent from both prompts at every base count. Two further reference-generated sets contain 32 uniform inputs and 32 inputs that activate both reference rules per family. The selected programs are executed on the same saved targets throughout the count extension. Binary success requires every example in a displayed or witness target to pass; broader-input agreement is the within-family fraction of correct outputs. When the fixed final-answer selector returns no legal program, task and new-input scores are set to zero; all reported agreement rates therefore measure the complete inference-and-delivery protocol rather than accuracy conditional on a valid returned program.

For any such family-level outcome $Y$, the paired effect at count $b$ is
\begin{equation}
\widehat\theta_b=\frac{1}{|F_b|}\sum_{f\in F_b}(Y_{fSb}-Y_{fCb}),\label{eq:effect}
\end{equation}
where $F_b$ is the certified family set. Bootstrap resampling preserves each family's paired observations \citep{efron1979,dror2018}. The original validation uses intervals adjusted across six model--target contrasts, and the original $H_2$ analysis adjusts across three model contrasts. The count extension reports pointwise 95\% intervals. Six planned comparisons of $\theta_b-\theta_5$ use the 119 families certified at every count and a Bonferroni adjustment. Appendices~\ref{app:prospective} and~\ref{app:r1completedpanel} specify the resampling procedures and analysis chronology.

\section{Results}\label{sec:results}
\subsection{Completion and witness agreement move in opposite directions}\label{sec:primary}
The original validation reveals a substantial completion decrease under certified separation (Table~\ref{tab:primary}). Qwen completes 67/120 control tasks and 12/120 separation tasks, a change of $-45.83\pp$. Olmo completes 52/120 and 15/120, a change of $-30.83\pp$. Their adjusted intervals exclude zero. Qwen has 55 control-only successes and no separation-only success; Olmo has 42 and five, respectively.

The same returned programs gain on the common-nine target: Qwen rises from zero to nine successes, and Olmo from zero to eleven. The gains are 7.50 and 9.17 percentage points, respectively, and both adjusted intervals exclude zero. On the wholly unshown pair, Qwen rises from zero to eleven successes and Olmo from zero to fourteen (Table~\ref{tab:h2}). Two of these successful separation programs from each model fail the displayed task. Thus the witness evaluation reveals behavior that complements task completion.

\begin{table}[t]
\centering{\small\setlength{\tabcolsep}{4pt}\input{tables/primary.tex}}
\caption{\textbf{Original five-base validation.} Displayed-task completion decreases for Qwen and Olmo while common-nine success increases. Intervals use the original adjustment across six planned contrasts. Phi uses the validated implementation described in Appendix~\ref{app:runtime}.}
\label{tab:primary}
\end{table}

\begin{table}[t]
\centering{\small\setlength{\tabcolsep}{4pt}\input{tables/h2.tex}}
\caption{\textbf{Success on the wholly unshown witness pair.} Both inputs are absent from both prompts. The original supplemental analysis uses intervals adjusted across the three model contrasts.}
\label{tab:h2}
\end{table}

Program-level execution makes the pattern concrete. Of Qwen's 67 displayed-correct controls, 60 pass the base and both control witnesses while failing both separation witnesses; Olmo has this pattern in 49 of 52. All 21 common-nine successes across the three models use rule sequences different from the reference sequence. Twenty contain two rules and one contains another legal number. The models therefore fit the finite target through alternative programs. Appendix~\ref{app:roles} gives their identities and witness-level outcomes.

\subsection{The pattern persists from three to six common examples}\label{sec:basecounts}
Table~\ref{tab:basecounts} reports the completed count extension. Every model--count cell has a negative displayed-task contrast and a positive wholly unshown-pair contrast. For Qwen, displayed changes at the new counts three, four, and six are $-36.97$, $-38.66$, and $-40.83\pp$. Olmo's corresponding changes are $-32.77$, $-23.53$, and $-34.17\pp$. Witness-pair gains range from 8.40 to 13.45 points across the full panel.

These results establish the observed direction across the tested example counts. The adjusted differences relative to base five range from $-3.36$ to $+8.40\pp$, with all six intervals spanning zero. Appendices~\ref{app:r1outcomes}, \ref{app:r1crosscount}, and~\ref{app:r1oneone} give the full-panel results, paired changes, and one-to-one observations.

Recertification also exposes the information carried by an unchanged example. One family's reference applies \texttt{f}$\to$\texttt{fdd}, followed by \texttt{db}$\to$\texttt{c}. Its base-five specification contains \texttt{iyxbdc}$\to$\texttt{iyxbdc}. Removing this identity constraint at base counts three and four admits the one-rule solution \texttt{b}$\to$\texttt{dc} in the reduced separation specification. That shortcut predicts \texttt{iyxdcdc} for the removed input. The example therefore excludes a concrete alternative even though its correct output repeats the input. The worked case in the anonymous supplement replays this change in the complete one-rule solution set.

\begin{table}[t]
\centering\small\setlength{\tabcolsep}{4pt}
\input{tables/basecounts_main.tex}
\caption{\textbf{Completion and witness success across common-example counts.} Each task displays $b+1$ examples. C/S gives displayed-task success counts. The two right columns give separation-minus-control changes and pointwise 95\% paired-family intervals, in percentage points.}
\label{tab:basecounts}
\end{table}

\subsection{Broader inputs reveal a persistent generalization challenge}\label{sec:broader}
The count extension executes all selected programs on the saved uniform and rule-active inputs. Uniform agreement falls in every model--count cell (Table~\ref{tab:broader}). This distribution contains many unchanged mappings: 3,325 of the 3,840 uniform inputs are identities under the reference. The uniform contrast therefore reflects the behavior of the returned programs on a largely identity-preserving distribution.

Rule-active inputs give positive point estimates throughout the panel. Five of eight pointwise intervals lie above zero, and the original base-five intervals span zero. Absolute separation agreement is 1.93--5.78\%. The combined pattern is consistent across counts: requiring a two-rule solution changes which witness examples the outputs satisfy, while broad rule-active accuracy remains low. Appendix~\ref{app:r1endpoints} gives the complete endpoint panel.

\begin{table}[t]
\centering\small\setlength{\tabcolsep}{4pt}
\input{tables/broader_main.tex}
\caption{\textbf{Behavior on separately sampled inputs.} UNIFORM agreement decreases throughout the panel. ACTIVE point estimates are positive throughout, with separation agreement below 6\%. Changes are separation minus control in percentage points, with pointwise 95\% paired-family intervals; the final column gives separation agreement in percent.}
\label{tab:broader}
\end{table}

\FloatBarrier
\subsection{Executable search and correct-candidate availability}\label{sec:robust}
A bounded symbolic solver enumerates one-rule programs first and then searches the two-rule candidates described in Appendix~\ref{app:solver}. It selects the first program satisfying the displayed task within ten CPU seconds. The solver completes all 120 original controls with one-rule programs, producing the witness failures predicted by Corollary~\ref{cor:short}. On separation, it completes 95/120 tasks and succeeds on the wholly unshown pair in 87/120 cases (Table~\ref{tab:solver}). This provides an executable reference point for the difficulty of solving the certified tasks.

\begin{table}[htbp]
\centering{\small\setlength{\tabcolsep}{4pt}\input{tables/solver.tex}}
\caption{\textbf{First-consistent symbolic search on the original tasks.} The solver searches from the displayed examples with a ten-CPU-second budget. The table tracks completion, shared-target success, and timeouts.}
\label{tab:solver}
\end{table}

The completion gap is already present in correct-candidate availability: Qwen expresses a displayed-task-correct candidate in 67 control and 13 separation trajectories, and Olmo in 57 and 17. Final successes are 67/12 and 52/15, respectively. Only one Qwen and two Olmo separation trajectories contain such a candidate without delivering a correct selected final program. Candidate availability and final-answer delivery are therefore distinct observable outcomes; these data do not attribute the main separation gap primarily to parsing or final delivery. Canonical and native extraction coincide for Qwen at 79/240 successes; Olmo has 67 canonical and four native successes, demonstrating sensitivity to accepted output notation. Appendix~\ref{app:implementation} gives the readout rules and complete counts.

Two further checks characterize the original outputs. Among the 46 families with identical prompt-token counts, Qwen and Olmo retain completion decreases of 45.65 and 26.09 points. Fixed token checkpoints show correct candidates and successful final answers emerging later in the saved trajectories. At 4,096 tokens, Qwen has 31/2 control/separation final successes, increasing to 67/12 at 8,192; Olmo increases from 6/1 to 52/15. Appendices~\ref{app:tokens} and~\ref{app:prefix} report the token and prefix measurements.

\subsection{Calibration of a saved task archive}\label{sec:archive}
Complete one-rule enumeration also provides a direct check on existing task records. In the saved 1,008-task archive, 23 tasks have certified minimum length one even though their references contain two or three rules. Another 226 have minimum length two, certified by one-rule exclusion and a two-rule reference. The remaining 759 have lower bound two and a reference-based upper bound of three, four, or five.

A concrete archive task illustrates the mismatch. Its three-rule reference replaces \texttt{v}$\to$\texttt{zi}, then \texttt{z}$\to$\texttt{zv}, then \texttt{z}$\to$\texttt{b}. A single replacement \texttt{v}$\to$\texttt{bvi} satisfies all five displayed pairs. On the additional input \texttt{z}, the two programs produce \texttt{bv} and \texttt{z}, respectively. Appendix~\ref{app:archiveexample} gives the full example and certificate counts; the supplement provides the archive records and executable checks. This audit shows how minimum-length checks can accompany an existing collection of executable examples.

\section{Conclusion}
Certificate-grounded calibration makes the shortest solution required by an example set directly verifiable. The archive audit reveals shorter solutions hidden by longer reference programs, and recertification shows how one unchanged example can exclude a shortcut. Across three to six shared examples, Qwen and Olmo complete fewer certified two-rule tasks while more often satisfying a wholly unshown witness pair. Correct-candidate availability falls alongside completion, and broad rule-active agreement remains low. These observations identify complementary evaluation targets for program search, final-answer delivery, and application to new inputs. The executable certificates and replayable outputs support further comparisons of program search, answer delivery, and behavior on new inputs under verified task requirements.

\section{Limitations}\label{sec:scope}
\paragraph{Task and certification scope.}
Our guarantees concern minimum program length in a bounded string-replacement language and a selected feeding-only reference population. The 120 families were selected from 15,004 proposals, and the example-count extension primarily varies identity constraints. The 23 shorter solutions characterize the finite 1,008-task saved collection, whose upstream sampling record is incomplete. Benchmark-wide prevalence estimation would require a documented sampling frame. Extending complete exclusion to longer programs or richer languages, and evaluating larger models, remain open directions.

\paragraph{Interpretation of the comparisons.}
The paired contrasts characterize a specification-level intervention that changes both the minimum required program length and the identifying information supplied by the distinguishing example. Witness tests measure agreement on construction-linked finite targets; global program equivalence and broad transfer require further evaluation. Agreement on rule-active inputs remains low. Family-bootstrap intervals describe variation across the selected families conditional on one sampled trajectory per model--family--condition--count cell. Replicated decoding seeds would quantify additional sampling variability. The six adjusted cross-count intervals span zero, leaving equivalence across example counts unresolved.

\paragraph{Computational scope.}
The symbolic baseline uses a restricted two-rule search and a finite CPU budget. A timeout or exhausted candidate pool leaves solution existence unresolved; nonexistence certification requires complete exclusion. Symbolic and neural systems use different computational resources. Prefix analyses characterize the progression of saved trajectories; forced stopping would be a separate intervention. The scope of the archived evidence and CPU replay is documented in Appendix~\ref{app:evidence}.
\label{main:end}

\section*{AI Use Statement}
Generative AI assisted hypothesis refinement, experimental design, implementation, proof checking, analysis, literature search, and writing. Executable generators produced the tasks and evaluation inputs. Numerical claims were checked against stored programs and family-level statistics. Figure~\ref{fig:construction} combines an AI-generated concept with an editable, semantically checked vector reconstruction. The author is responsible for all claims, references, and code.

\section*{Reproducibility Statement}
The companion research artifact provides the task families, example-count variants, certificate records, evaluation inputs, normalized Qwen3-8B and Olmo-3-7B-Think outputs for the 1,920-trajectory panel, and the corresponding family-level results. Its CPU replay rechecks the certificates, evaluates selected programs, and reconstructs the count-panel statistics without model inference.

The versioned artifact, file manifest, and reproduction instructions are available at\par
\noindent\url{https://github.com/dqswordman/program-structure-certificates}\par
Appendix~\ref{app:evidence} distinguishes directly replayable results from earlier analyses documented through aggregate tables and protocols.

\section*{Ethics Statement}
The study uses generated string-transformation tasks and released models, with no human participants or personal records.

\setlength{\bibsep}{3pt plus 1pt minus 1pt}
\bibliography{references}
\bibliographystyle{plainnat}
\clearpage
\appendix
\renewcommand{\theHfigure}{appendix.\thesection.\arabic{figure}}
\renewcommand{\theHtable}{appendix.\thesection.\arabic{table}}
\input{appendix.tex}
\input{r1_certificate_audit.tex}
\end{document}

%% file: tables/primary.tex
\begin{tabular}{llrrrr}
\toprule
Model & Target & Control & Separation & $S-C$ (pp) & Adjusted interval\\
\midrule
Qwen & Visible six & 67/120 & 12/120 & -45.83 & [-57.50, -34.17]\\
 & Common nine & 0/120 & 9/120 & +7.50 & [+1.67, +14.17]\\
\addlinespace[2pt]
Olmo & Visible six & 52/120 & 15/120 & -30.83 & [-44.17, -17.50]\\
 & Common nine & 0/120 & 11/120 & +9.17 & [+3.33, +16.67]\\
\addlinespace[2pt]
Phi$^{\dagger}$ & Visible six & 5/120 & 2/120 & -2.50 & [-8.33, +3.33]\\
 & Common nine & 0/120 & 1/120 & +0.83 & [+0.00, +3.33]\\
\bottomrule
\end{tabular}

%% file: tables/h2.tex
\begin{tabular}{lrrrr}
\toprule
Model & Control & Separation & $S-C$ (pp) & Adjusted interval\\
\midrule
Qwen3-8B & 0/120 & 11/120 & +9.17 & [+3.33, +15.83]\\
Olmo-3-7B-Think & 0/120 & 14/120 & +11.67 & [+5.00, +19.17]\\
Phi-4-mini$^{\dagger}$ & 1/120 & 1/120 & +0.00 & [-2.50, +3.33]\\
\bottomrule
\end{tabular}

%% file: tables/basecounts_main.tex
\begin{tabular}{llrrrr}
\toprule
Model & $b$ & $N$ & C/S & Displayed $\Delta$ [CI] & Unshown pair $\Delta$ [CI]\\
\midrule
Qwen & 3 & 119 & 59/15 & $-36.97\, [-46.22,-27.73]$ & $+13.45\, [+7.56,+20.17]$\\
 & 4 & 119 & 61/15 & $-38.66\, [-47.90,-28.57]$ & $+10.92\, [+5.88,+16.81]$\\
 & 5 & 120 & 67/12 & $-45.83\, [-55.00,-36.67]$ & $+9.17\, [+4.17,+15.00]$\\
 & 6 & 120 & 68/19 & $-40.83\, [-50.00,-31.67]$ & $+12.50\, [+6.67,+18.33]$\\
\addlinespace[3pt]
Olmo & 3 & 119 & 53/14 & $-32.77\, [-42.86,-22.69]$ & $+8.40\, [+3.36,+13.45]$\\
 & 4 & 119 & 47/19 & $-23.53\, [-34.45,-12.61]$ & $+11.76\, [+5.88,+17.65]$\\
 & 5 & 120 & 52/15 & $-30.83\, [-40.83,-20.83]$ & $+11.67\, [+6.67,+17.50]$\\
 & 6 & 120 & 54/13 & $-34.17\, [-43.33,-25.00]$ & $+9.17\, [+4.17,+15.00]$\\
\bottomrule
\end{tabular}

%% file: tables/broader_main.tex
\begin{tabular}{llrrr}
\toprule
Model & $b$ & UNIFORM $\Delta$ [CI] & ACTIVE $\Delta$ [CI] & ACTIVE S (\%)\\
\midrule
Qwen & 3 & $-23.19\, [-30.75,-15.49]$ & $+5.30\, [+1.71,+9.45]$ & 5.78\\
 & 4 & $-27.26\, [-34.59,-20.01]$ & $+3.39\, [+0.10,+7.22]$ & 4.33\\
 & 5 & $-36.12\, [-43.88,-28.33]$ & $+2.16\, [-0.29,+5.29]$ & 2.76\\
 & 6 & $-30.96\, [-38.88,-22.99]$ & $+4.53\, [+1.20,+8.59]$ & 5.00\\
\addlinespace[2pt]
Olmo & 3 & $-23.45\, [-31.88,-14.97]$ & $+4.67\, [+1.18,+8.74]$ & 5.28\\
 & 4 & $-18.46\, [-27.26,-9.74]$ & $+4.49\, [+1.26,+8.25]$ & 4.94\\
 & 5 & $-21.72\, [-29.58,-13.75]$ & $+1.59\, [-0.34,+4.22]$ & 2.01\\
 & 6 & $-27.92\, [-35.55,-20.36]$ & $+1.46\, [-0.49,+4.19]$ & 1.93\\
\addlinespace[2pt]
\bottomrule
\end{tabular}

%% file: tables/solver.tex
\begin{tabular}{lrrrr}
\toprule
Condition & Visible six & Common nine & Unshown $H_2$ & Timeouts\\
\midrule
Control & 120/120 & 0/120 & 0/120 & 0\\
Separation & 95/120 & 86/120 & 87/120 & 25\\
\bottomrule
\end{tabular}

%% file: appendix.tex
\FloatBarrier
\section{Construction, Certificates, and Sampling}\label{app:construction}
\subsection{Reproducible enumeration}
For $n$ examples, let $u=|A(E)|$ and $v=|B(E)|$. The one-rule solver performs at most $uvn$ replacement checks. Lemma~\ref{lem:complete} establishes completeness of the example-derived candidate strings. Find strings have positive length, and replacement candidates include the empty string. We deduplicate repeated input and output substrings. Each certificate records the complete one-rule solution set, the two-rule reference, and the example specification.

We generate candidate witness strings from length-two/three strings over the alphabet, base/rule anchors, and deterministic repeated or sampled contexts around those anchors, retaining lengths two through six. The fixed reference produces their outputs. For each five-feature signature and condition, a fixed hash ordering retains up to four representatives. Deterministic selection then chooses two disjoint control/separation pairs with four distinct witness inputs. This construction defines the eligible sampling population.

Filtering the complete base solution set determines witness eligibility. Independent re-enumeration verifies all 480 six-example certificates and 240 strict matches.

\subsection{Complete example}
The reference is the ordered program
\begin{align*}
 &\operatorname{replace}(\texttt{ex},\texttt{azw}),\\
 &\operatorname{replace}(\texttt{wa},\texttt{y}).
\end{align*}
The following nine mappings are the five base examples and both witness pairs:
\begin{center}\small
\begin{tabular}{@{}lll@{}}
\toprule
Role & Input & Output\\
\midrule
Base & \texttt{yf} & \texttt{yf}\\
Base & \texttt{gud} & \texttt{gud}\\
Base & \texttt{czua} & \texttt{czua}\\
Base & \texttt{dzexa} & \texttt{dzazy}\\
Base & \texttt{bwcied} & \texttt{bwcied}\\
Control 1 & \texttt{iiiexa} & \texttt{iiiazy}\\
Separation 1 & \texttt{wbawex} & \texttt{wbayzw}\\
Control 2 & \texttt{hhexa} & \texttt{hhazy}\\
Separation 2 & \texttt{azwex} & \texttt{azyzw}\\
\bottomrule
\end{tabular}
\end{center}
The base has a unique one-rule solution, \texttt{exa}$\to$\texttt{azy}. Both control specifications preserve it. Complete enumeration of each separation specification yields an empty one-rule solution set, and the two-rule reference satisfies all nine mappings. The two matched signatures are $(6,6,3,3,0)$ and $(5,5,3,2,0)$. This expository family comes from the pre-inference design record.

\subsection{Eligibility and sampling population}
The validation generator records 15,004 proposals: 11,970 outside the desired reference-relation class; 1,404 with an empty one-rule base solution set; 1,169 excluded by prior identity; 338 lacking two disjoint strictly matched pairs; three identity cases; and 120 accepted families. Among accepted bases, 119 have one nonidentity mapping and one has two. Within the one-to-five-rule language, complete one-rule enumeration gives $|\Sone(B_f)|=1,2,3$ in 91, 13, and 16 families, respectively (minimum/median/maximum $1/1/3$).

The reference key normalizes the program representation, including its recorded symbol mapping. Exclusion checks cover the archived earlier references, proposals, parsed model programs, and constructive alternatives. The exclusion operates on normalized program syntax. The reported family bootstrap treats the accepted families as the independent sampling units in this generator-conditioned design.

A constructive search finds nonfeeding minimum-length alternatives for 152 separation specifications. The remaining 88 cases have unresolved feeding-dependence status.

\FloatBarrier
\section{Discovery Evidence and Readout Sensitivity}\label{app:discovery}
Discovery 1 used 60 families, two models, two specification conditions, and two instruction versions, yielding 480 outputs. Its primary Qwen completion contrast was $-32.50$ points. Of these, 22 families met all five strict surface matches. The Olmo readout is sensitive to rule notation: the strict detector counted 11/120 correct-candidate trajectories under the original instruction and 35/120 under the explicit instruction. Canonicalizing the predefined rule notations changed those counts to 30/120 and 35/120, reducing the apparent instruction effect from 20.00 to 4.17 points.

Discovery 2 used 60 new, strictly matched families, two witness pairs, and the fixed explicit instruction. Its 480 outputs gave canonical visible-six effects of $-26.67$ points for Qwen and $-15.83$ for Olmo. Both witness-specific point estimates were negative. An exploratory common-nine analysis gave gains of $+21.67$ and $+15.00$ points.

Before generating validation outputs, we fixed the common target, canonical parser, witness assignment, and six primary contrasts. The summaries above preserve the discovery observations and native-readout sensitivity; Appendix~\ref{app:evidence} identifies their documentation and replay coverage.

\FloatBarrier
\section{Inference Configuration and Final-Answer Rules}\label{app:implementation}
\subsection{Pinned inference configuration}
The exact model repositories and revisions are:
\begin{quote}\footnotesize\raggedright
\texttt{Qwen/Qwen3-8B}\\
\nolinkurl{b968826d9c46dd6066d109eabc6255188de91218}\\[3pt]
\texttt{allenai/Olmo-3-7B-Think}\\
\nolinkurl{d97e442d7cc678210054dbcc9b440894d62c89a4}\\[3pt]
\texttt{microsoft/Phi-4-mini-reasoning}\\
\nolinkurl{0e3b1e2d02ee478a3743abe3f629e9c0cb722e0a}
\end{quote}
The environment uses PyTorch 2.8.0+cu128 and Transformers 4.57.1. Generation uses BF16 weights, SDPA, sampling temperature 0.6, top-$p$ 0.95, one return sequence, and repetition penalty 1.0. Top-$k$ is 20 for Qwen and zero (disabled) for Olmo and Phi. Batch sizes are eight, four, and four. Model-native EOS tokens are respected. The generation budget is 8,192 new tokens for every model.

All model evaluations use an RTX 5090.

For the original six-example validation, maximum rendered prompt lengths are 272, 339, and 281 tokens for Qwen, Olmo, and Phi. Model revisions, token counts, and random seeds identify the recorded evaluations. The companion artifact records task specifications, selected programs, the verbatim synthesis instruction, native input examples, and the pinned Qwen/Olmo chat templates. We record the surface features and model-native token counts for each pair. The original bootstrap used seed \texttt{12645822550545217882} and an archived $50{,}000\times120$ index array.

\subsection{Verbatim synthesis instruction and native inputs}\label{app:verbatim-prompts}
Each original validation input uses the following instruction, followed by six input--output pairs. The notation example demonstrates the output serialization.
\begin{Verbatim}
Infer a cascade of string-replacement rules that maps every input to its paired output.
Apply each rule to every string in order, using Python str.replace semantics.
Use only characters from abcdefghijkuvwxyz.
The first argument must contain 1 to 3 characters; the second may contain 0 to 3 characters.
Use at most 5 rules. Any functionally correct cascade is acceptable.
You may reason before answering. Put your final answer in one fenced block labelled python.
Inside that block write a Python list whose elements are strings. Each string must be a rule written as replace('a','b'), with two quoted arguments satisfying the limits above.
For example, the serialization of one rule is:
```python
["replace('a','b')"]
```
This example shows notation only, not a proposed solution. Keep the order of your rules. Arrow notation, bare character pairs, and a list of output strings are not the requested rule-list format.
Input-output pairs:
\end{Verbatim}

For the family in Appendix~\ref{app:construction}, the recorded display order is \texttt{gud}, \texttt{czua}, \texttt{bwcied}, \texttt{dzexa}, \texttt{yf}, followed by \texttt{iiiexa}$\to$\texttt{iiiazy} in control or \texttt{wbawex}$\to$\texttt{wbayzw} in separation. The two conditions contain 265 tokens for Qwen, 332 for Olmo, and 272 for Phi. These counts include native framing. The other witness pair is reserved for evaluation.

The native templates add model-specific framing to a single user message: Qwen uses its user/assistant delimiters with thinking enabled; Olmo inserts its default system message and opening reasoning tag; Phi inserts its default system message. The Qwen/Olmo panel in the companion artifact records prompt token arrays, user-message content, exact chat templates, and pinned model/tokenizer metadata, together with rendered prompt strings for the new trajectories. Qwen's template is an export of its tokenizer-configuration field; Olmo's template is the archived standalone file. Appendix~\ref{app:available-artifacts} identifies the included records, and Appendix~\ref{app:replay-scope} identifies analyses retained as protocols and aggregate tables.

\subsection{Selection precedes execution}
The canonical selector chooses a program from the text before executing evaluation targets. For Qwen and Olmo, the eligible segment begins after a closed thinking delimiter. The selector takes the last complete Python fence, falling back to the last complete list when a complete Python fence is absent. Finite, frozen notation mappings preserve find/replacement strings and rule order. Final success equals one when the selected object is a legal program that satisfies the target, and zero otherwise.

Phi's frozen adapter accepts either the same closed-thinking route or an eligible terminal block/list at actual EOS when no explicit unfinished thinking segment is present. A fixed terminal-token set defines eligible suffixes; an unfinished reasoning segment remains ineligible. Evaluation protocol validation includes 19 parsing cases for Qwen and Olmo and 33 model-boundary cases.

Native and canonical scoring use their specified endpoint eligibility rules. The re-evaluated Phi cohort has nine native successes and seven canonical successes. Candidate availability scans the trajectory for an admissible program that satisfies the target; parseability records successful extraction of a legal final program. These readouts track expressed candidates, delivered programs, and their execution outcomes.

The recorded fields include the selected span, legal rule sequence, eight executable target scores, all three readout layers, generation length, and EOS/CAP status. Each original model--condition cell contains 120 outputs. Prefix observations at 2,048, 4,096, and 6,144 generated tokens provide nested measurements of those trajectories.

\begin{table}[t]
\centering{\small\setlength{\tabcolsep}{4pt}\input{tables/readout.tex}}
\caption{Stopping, extracted programs, and successful execution in the original validation. $^{\dagger}$ marks the corrected Phi implementation.}
\label{tab:readout}
\end{table}

At the full budget, Qwen expresses a correct candidate in 13 separation trajectories and delivers a correct final program in 12; Olmo's corresponding counts are 17 and 15. Thus one Qwen trajectory and two Olmo trajectories contain a correct candidate and receive zero final success under the fixed readout. Control candidate/final counts are 67/67 for Qwen and 57/52 for Olmo. These counts quantify the observed gap between candidate expression and successful final delivery.

\FloatBarrier
\section{Attention validation and uniform re-evaluation}\label{app:runtime}
\subsection{Cohort definition}
This appendix documents the original Phi evaluation and numerical checks. The companion artifact retains these summaries and protocols; original Phi trajectories and detailed attention-validation logs are absent (Appendix~\ref{app:replay-scope}).

All reported Phi results use one corrected causal-attention implementation. Correction of the causal-attention position and mask specification was followed by uniform re-evaluation of the entire 240-output cohort. Re-evaluation retained the original tasks, prompts, token IDs, weights, sampling parameters, seeds, batch size, token budget, and readout; every result uses this common implementation. The 480 Qwen/Olmo outputs are unchanged. The cohort retains the original task sample.

\subsection{Numerical checks}
The corrected implementation was fixed before the full Phi cohort was re-evaluated. Independent causal-plus-padding masks and FP32 attention computations provide the numerical references; the tested full-prefix/reset logits and cache entries match the independent-mask reference bitwise. Full-versus-incremental BF16 comparisons give maximum absolute logit difference 0.3125, maximum mean logit difference 0.05849, and maximum absolute cache difference 0.25, with identical tested argmax predictions. The reported validation summary above accompanies the common implementation used for the original Phi results.

\subsection{Stopping and readout outcomes}
The corrected Phi cohort contains five/two visible successes, zero/one common-audit successes, and 13/eight parseable finals under control/separation. EOS occurs on 23/20 trajectories and the token cap on 97/100. Among all 240 outputs, 195 reach the cap without a final boundary, one has a capped unparseable final, one a capped incorrect final, 23 end at EOS without a parseable final, 13 end with an incorrect final, and seven provide a successful final.

\FloatBarrier
\section{Statistical Estimation and Analysis Provenance}\label{app:prospective}
\subsection{Paired estimands and uncertainty}\label{app:paired-uncertainty}
Each accepted family is the resampling unit, and the separation-minus-control contrast is computed on its paired outcomes as in Eq.~\eqref{eq:effect}. For the original validation, six planned model--target contrasts use 50,000 family-bootstrap resamples, seed \texttt{12645822550545217882}, and percentile quantiles $1/240,239/240$. The supplemental $H_2$ analysis uses the same original resampling schedule with quantiles $1/120,119/120$ for its separate three-contrast adjustment. The example-count extension uses 20,000 paired-family resamples with PCG64 seed \texttt{20261107} and pointwise 95\% intervals. Its six planned displayed-success comparisons against base five use quantiles $1/240,239/240$ on the fixed 119-family common-certificate cohort. The adjustments apply to their specified comparison families; remaining secondary endpoints use descriptive estimates or pointwise intervals.

Marginal 95\% intervals for original visible/common differences are $[-55.00,-37.50]$/$[3.33,12.50]$ for Qwen, $[-40.83,-20.83]$/$[4.17,15.00]$ for Olmo, and $[-6.67,1.67]$/$[0,2.50]$ for corrected Phi. Marginal $H_2$ intervals are $[4.17,15.00]$, $[6.67,17.50]$, and $[-2.50,2.50]$ points, respectively. Tables~\ref{tab:primary} and~\ref{tab:h2} retain their original adjusted intervals. Equal-token and token-difference strata were analyzed using recorded subset family IDs and 50,000 paired-family resamples; strata with $n<12$ are descriptive diagnostics. Appendix~\ref{app:replay-scope} distinguishes these historical summaries from the replayable count-panel resampling.

\subsection{Exact paired sensitivity}\label{app:exact-sensitivity}
To check sensitivity of the six original binary contrasts to the bootstrap method, we also report two-sided exact paired-binary tests with a six-comparison Bonferroni correction. Let $g$ and $\ell$ count separation-only and control-only successes. Under independent paired sampling and paired marginal symmetry, $g\mid(g+\ell)$ is $\operatorname{Binomial}(g+\ell,1/2)$. We use a two-sided exact tail test and $\min(1,6p)$ \citep{mcnemar1947,dror2018}. These tests complement the planned percentile bootstrap.

\begin{table}[htbp]
\centering{\small\setlength{\tabcolsep}{4pt}\input{tables/exact.tex}}
\caption{Exact paired-binary tests for the six primary contrasts. Gain and loss denote separation-only and control-only successes.}
\label{tab:exact}
\end{table}

For common-nine, Qwen has nine gains and zero losses, giving $p=2/2^9=.00390625$ and corrected $p=.0234375$. Olmo has eleven gains and zero losses, giving $p=.0009765625$ and corrected $p=.005859375$. Phi has one gain. This supplemental analysis supports the reported direction of the Qwen and Olmo common-nine comparisons, while providing no corresponding positive conclusion for Phi.

\FloatBarrier
\subsection{Analysis chronology and original magnitude benchmark}\label{app:analysis-history}
The two discovery studies preceded the independent original validation. The original common-nine target, final-answer selector, witness assignment, and six primary contrasts were fixed before that validation's model generation. The isolated wholly unshown $H_2$ contrast and other supplemental diagnostics were specified after the original outputs were generated but before their supplemental scoring. These diagnostics include broader-input generation, symbolic-solver order, alternative selectors, solution-count groups, token strata, and prefix measurements. The subsequent 3/4/6-base panel uses new model trajectories and is reported as a separate example-count extension. The original analysis plan also included a ten-percentage-point criterion for a joint magnitude interpretation; the observed common-nine gains of 7.50 and 9.17 points did not meet that criterion. The companion artifact preserves the full historical rule and its original manuscript wording in \texttt{provenance/ANALYSIS\_HISTORY.json}.

\FloatBarrier
\section{Program Behavior on Shared Witnesses}\label{app:roles}
The fixed nine-role order is $(B_1,\ldots,B_5,c_1,s_1,c_2,s_2)$. We execute the selected final program on all nine mappings and record a binary score for each. A missing program receives an all-zero vector. For displayed index $j$, the supplemental target is exactly $(c_{3-j},s_{3-j})$, the shared pair held out from both prompts. We also record success on the base-five, both-control, and both-separation sets.

For Qwen, 60 of 67 visible-correct controls have bit vector $(1,1,1,1,1,1,0,1,0)$, passing the shared base and both control witnesses while failing both separation witnesses. Six have $(1,1,1,1,1,1,0,0,0)$ and one has $(1,1,1,1,1,0,0,1,0)$. Olmo's corresponding counts are 49, two, and one out of 52. All five visible-correct Phi controls pass the base and control witnesses and fail the two separation witnesses. These records locate the errors within the 120-family evaluation.

Qwen has two additional $H_2$-successful separation outputs beyond its nine common successes, and both fail the visible specification. Olmo has three additional $H_2$ successes: two fail visible six, and one passes visible six and fails common nine. Phi's one control $H_2$ success fails visible six.

\begin{table}[htbp]
\centering{\small\setlength{\tabcolsep}{4pt}\input{tables/program_classes.tex}}
\caption{Program lengths and ordered-rule identities among successful witness-target predictions.}
\label{tab:programs}
\end{table}

Rule-sequence equality compares normalized ordered $(a,b)$ pairs. In Table~\ref{tab:programs}, ``Reference'' denotes exact sequence equality and ``Other 2'' denotes a different two-rule sequence. Every common-nine-successful output uses a sequence different from the reference.

Corollary~\ref{cor:short} follows directly from the construction. Every one-rule control solution is in $\Sone(B_f)$; each $s_i$ excludes all such programs. We independently re-enumerated all 480 six-example specifications and checked this implication in all 120 families. The same references pass every witness, demonstrating that the prediction depends on the returned program class.

\subsection{Behavior by base solution count}\label{app:groups}
The base certificates define two groups: $|\Sone(B_f)|=1$ (91 families) and $|\Sone(B_f)|>1$ (29). Table~\ref{tab:solution-groups} reports visible-six, common-nine, and $H_2$ outcomes for every model. C and S denote success counts among the $n$ families in each group. Both Qwen and Olmo have lower visible-six success and higher common-nine success under separation in each group. The completion decrease and common-target gain are larger in the multiple-solution group. We compute marginal, descriptive 95\% intervals with 50,000 paired-family resamples within each group. Phi uses the common validation implementation.

\begin{table}[htbp]\centering\small\setlength{\tabcolsep}{3.2pt}
\input{tables/solution_groups}
\caption{\textbf{Paired behavior by the number of one-rule base solutions.} The two groups summarize tasks whose bases admit one or multiple short solutions.}
\label{tab:solution-groups}
\end{table}

The group analysis executes the selected programs on the fixed scoring targets and resamples families while keeping both conditions paired. In the multiple-solution Phi group, every common-nine and $H_2$ paired difference is zero, yielding empirical intervals $[0,0]$. The original analysis record identifies the full-sample and group statistics, family memberships, and resampling indices.

\FloatBarrier
\section{Prompt Tokenization Diagnostics}\label{app:tokens}
Counts use the archived model-native prompt token IDs, including template overhead. A paired delta is separation length minus control length. Qwen and Olmo each have 46 equal-count families (38.33\%); Phi has 55 (45.83\%). Summary statistics appear in Table~\ref{tab:tokens}, and Table~\ref{tab:tokenstrata} gives the exact stratum counts and completion contrasts. For the five predefined token-difference strata, we report marginal intervals from paired-family resampling. Strata with fewer than 12 families provide descriptive diagnostics.

\begin{table}[htbp]
\centering{\small\setlength{\tabcolsep}{4pt}\input{tables/token.tex}}
\caption{Prompt-token differences within the matched specification pairs.}
\label{tab:tokens}
\end{table}

\begin{table}[htbp]
\centering{\small\setlength{\tabcolsep}{4pt}\input{tables/token_strata.tex}}
\caption{Displayed-task completion differences across the five predefined prompt-token strata.}
\label{tab:tokenstrata}
\end{table}

The zero-delta subset retains large negative visible effects. Its common-nine changes are $+6.52$ points ($[0,15.22]$ marginal) for Qwen and $+13.04$ ($[4.35,23.91]$) for Olmo; its $H_2$ changes are $+10.87$ ($[2.17,19.57]$) and $+17.39$ ($[6.52,28.26]$). Phi has one visible success in each zero-delta condition. These subsets quantify the comparison at equal prompt length.

\FloatBarrier
\section{Program Behavior on Broader Inputs}\label{app:extra}
\subsection{Generation and metrics}
For each family, we generate 32 uniform and 32 active inputs, using distinct strings outside the original nine inputs. Uniform inputs use the same 17-character alphabet with lengths two through twelve. Active proposals combine fixed reference-rule anchors with sampled contexts. Acceptance requires both reference replacements to change their respective intermediate strings. Reference execution determines the activation criterion. Seeds, proposal counts, rejections, and accepted inputs were fixed before supplemental scoring. Each layer has a 10,000-proposal ceiling per family; all requested inputs were obtained.

For program $\hat p_{fc}$ and input stratum $A_f$ of size 32, we report
\[
Z_{fc}=\frac1{32}\sum_{x\in A_f}\ind[\hat p_{fc}(x)=p_f^*(x)],
\]
We set $Z_{fc}=0$ when no final program is selected. The reported contrast is the mean of $Z_{fS}-Z_{fC}$ over 120 families; the machine-readable results also contain all-pass rates. The 32 inputs per stratum give 3,840 input tests per layer and 7,680 overall. Both conditions use the same input sets, generated from the reference program. We resample paired families to obtain marginal intervals for each of the six model--stratum combinations in Table~\ref{tab:extra}. Symbolic-solver timeouts also receive $Z_{fc}=0$.

\begin{table}[htbp]
\centering{\small\setlength{\tabcolsep}{4pt}\input{tables/extra.tex}}
\caption{Program accuracy on uniformly sampled and rule-active inputs.}
\label{tab:extra}
\end{table}

\subsection{Accuracy across input distributions}
Of the 3,840 uniform inputs, 3,325 (86.59\%) are unchanged by the reference. Every active input has a nonidentity reference output. Uniform accuracy therefore reflects identity behavior alongside the availability of a returned program. Qwen and Olmo have positive active-layer point estimates, with intervals spanning zero.

Executing the frozen symbolic outputs gives uniform mean accuracies of 87.19\%/70.26\% and active means of 0.42\%/40.42\% under control/separation. These programs come from the original search budget. Their accuracy profile, together with Table~\ref{tab:extra}, shows how returned programs behave across the two input distributions.

\FloatBarrier
\section{Bounded Symbolic Search}\label{app:solver}
\subsection{Solver inputs and search space}
Each solver task contains an opaque ID, the legal language specification, and the ordered six visible pairs. The search commits its selected program before an external evaluator executes the reference-defined targets. The evaluator holds the certificates, unshown witnesses, audit inputs, and recorded model outputs. The public string-replacement semantics define the search language.

Candidate rules are ordered by $(|a|+|b|,|a|,a,b)$ after deterministic deduplication. One-rule enumeration is complete as in Lemma~\ref{lem:complete}. For the first rule of a two-rule program, find strings come from visible input substrings of length at most three; replacements come from visible input/output substrings, public alphabet singletons, and the empty string. Each candidate produces an intermediate input vector. An unchanged vector is skipped. The second rule is enumerated completely from that intermediate vector to the visible outputs, using the one-rule lemma. The search returns and commits the first consistent program. Complete one-rule enumeration supplies the certificate; the restricted two-rule pool supplies a search baseline whose timeouts or exhausted pools leave existence unresolved.

\begin{algorithm}[htbp]
\caption{First-consistent bounded solving of a displayed specification}
\begin{tabularx}{\linewidth}{@{}r@{\quad}>{\raggedright\arraybackslash}X@{}}
&\textbf{Input:} Visible specification $E$; public language; CPU limit $\tau$\\
1&Start timer before candidate construction.\\
2&\textbf{for} one-rule candidate $r$ in the complete, fixed order:\\
3&\quad Record elapsed-time/candidate checkpoints; stop if limit reached.\\
4&\quad\textbf{if} $r\models E$, \textbf{return} $[r]$.\\
5&\textbf{end for}\\
6&\textbf{for} first rule $r_1$ in the fixed restricted pool:\\
7&\quad Compute intermediate inputs $z_i=r_1(x_i)$; skip an unchanged vector.\\
8&\quad\textbf{for} second rule $r_2$ in the complete one-rule search from $z_i$ to $y_i$:\\
9&\qquad Record checkpoints; stop if limit reached.\\
10&\qquad\textbf{if} $[r_1,r_2]\models E$, \textbf{return} $[r_1,r_2]$.\\
11&\quad\textbf{end for}\\
12&\textbf{end for}\\
13&\textbf{return} no selected program.\\
\end{tabularx}
\end{algorithm}

\subsection{Budgets and recorded execution}
Each of the 240 tasks runs once with a ten-process-CPU-second cap and a 60-second wall-clock safeguard. Each condition contains 120 tasks. The same search records checkpoints at 0.1, one, and ten CPU seconds and at 1,000, 10,000, and 100,000 full candidate-program checks. A selected program is retained at later checkpoints. CPU time includes candidate construction and search; the record reports process startup separately. One candidate check tests a full program against the visible specification, allowing early rejection. The record also contains per-example execution counts.

\begin{table}[htbp]
\centering{\small\setlength{\tabcolsep}{4pt}\input{tables/solver_anytime.tex}}
\caption{Success across CPU-time checkpoints of the bounded symbolic search. The ten-second checkpoint gives the primary result.}
\label{tab:solveranytime}
\end{table}
\begin{table}[htbp]
\centering{\small\setlength{\tabcolsep}{4pt}\input{tables/solver_candidates.tex}}
\caption{Success across full-program candidate-check checkpoints of the same symbolic runs.}
\label{tab:solvercandidates}
\end{table}

The primary run solves 215 tasks, with 25 reaching the CPU-time limit. It uses 10.47 CPU-minutes in total and approximately $1.924\times10^8$ candidate checks, with peak working-set memory of 23.7 MiB. All 120 control selections contain one rule; all 95 separation selections contain two.

A repeated execution agrees on 238/240 final selections. The two differences occur at the time limit; the 215 tasks solved in both executions have identical programs and candidate counts. Tables~\ref{tab:solveranytime}--\ref{tab:solvercandidates} use the original run throughout. Fixed candidate-order checkpoints complement the machine-dependent time-budget measurements.

\FloatBarrier
\section{Candidates and Final Answers Across Token Prefixes}\label{app:prefix}
\subsection{Nested trajectory measurements}
The 720 original validation trajectories provide four nested prefixes at 2,048, 4,096, 6,144, and 8,192 generated tokens, totaling 2,880 records. A trajectory ending before a cut retains its actual EOS; otherwise the prefix is marked capped. We decode text with the archived tokenizer and original boundary contract and record parseability, visible/common scores, and correct-candidate availability at every cut. Each checkpoint includes the 120 trajectories in each model--condition cell, with zero final success assigned when an eligible final program is absent. Phi contributes the same 240 outputs at each cut. These measurements describe saved trajectories rather than a separate forced-stopping intervention.

\begin{table}[htbp]
\centering{\small\setlength{\tabcolsep}{4pt}\input{tables/prefix.tex}}
\caption{Parseable finals, displayed-task success, and correct-candidate availability across recorded token prefixes.}
\label{tab:prefix}
\end{table}

At 2,048 tokens, Qwen expresses ten correct control candidates and one correct separation candidate, while Olmo expresses two correct control candidates. Canonical final success is zero for both models at this checkpoint. The later checkpoints trace the appearance of final answers within the recorded 8,192-token budget.

\subsection{Two fixed alternative selectors}
For Qwen and Olmo, alternative A selects the first complete Python fence in the eligible final segment, or the first complete list when a fence is absent. Alternative B selects the last complete object by character endpoint across fences and standalone lists, deduplicating lists contained within a fence. Both use the frozen final-segment eligibility rule and legal-program parser. Object selection precedes parsing and execution; a selected object outside the grammar receives zero final success.

Across 480 outputs and two alternatives, all 960 selected rule sequences and their visible-six, common-nine, and $H_2$ scores agree with the primary selection. Phi uses its frozen endpoint rule.

\FloatBarrier
\section{Artifact Inventory and Reproduction Scope}\label{app:evidence}
\subsection{Available artifacts}\label{app:available-artifacts}
The companion artifact is organized around three independently checkable objects: (i) task specifications and complete one-rule certificates; (ii) saved Qwen/Olmo model outputs and the deterministic final-program readout; and (iii) evaluation targets and family-level statistics. It includes the 120-family example-count variants, 960 arm-specific certificates, 1,008 saved archive tasks, 7,680 additional reference-generated targets, and normalized records for the 1,920-trajectory Qwen/Olmo panel. The provided CPU entry point verifies the certificates, re-executes selected programs, and reconstructs the recorded count-panel statistics without rerunning model inference.

The following file map identifies the corresponding objects; paths are relative to the repository root.
\begin{center}\small
\begin{tabularx}{\linewidth}{@{}>{\raggedright\arraybackslash}p{.43\linewidth}>{\raggedright\arraybackslash}X@{}}
\toprule
Files & Contents and role\\
\midrule
\nolinkurl{data/families.jsonl}, \nolinkurl{data/base_variants.jsonl}, \nolinkurl{data/certificate_records.jsonl} & 120 families, 480 paired variants, and 960 arm-specific certificates.\\[3pt]
\nolinkurl{data/native1008.csv}, \nolinkurl{data/broader_inputs.jsonl} & Saved archive examples and the 7,680 broader-input targets.\\[3pt]
\nolinkurl{data/trajectories.jsonl.gz}, \nolinkurl{data/model_contract.json} & Normalized outputs, token IDs, selected programs, and pinned Qwen/Olmo configuration.\\[3pt]
\nolinkurl{results/}, \nolinkurl{reproduce.py} & Recorded scientific tables and the CPU reconstruction entry point.\\[3pt]
\nolinkurl{MANIFEST.json}, \nolinkurl{RELEASE_SCOPE.txt} & File hashes, availability status, replay commands, and result-level coverage.\\
\bottomrule
\end{tabularx}
\end{center}

The replay reconstructs the 1,920-row per-trajectory metric table, 348-row paired-effect table, and 2,872-row family-difference table, including the count-panel bootstrap intervals. All 1,920 records retain prompt and generated token IDs; the 1,440 new records also retain complete rendered prompt text. The 480 base-five records provide reconstructed user-message content, prompt hashes, and the pinned templates. A small end-to-end fixture connects a displayed specification, complete one-rule enumeration, a saved selected program, and execution on displayed, witness, uniform, and active inputs. From the package root, run:
\begin{quote}\small
\texttt{python -B verify\_manifest.py}\\
\texttt{python -B replay\_fixture.py}\\
\texttt{python -B reproduce.py --out replay}
\end{quote}
The output directory for the full replay must be new. The manifest records file identities, cohort and model coverage, availability status, replay entry points, and reproduced quantities. Its separate unavailable-record entries identify historical results represented by summaries and protocols.

\subsection{Execution scope and historical records}\label{app:replay-scope}
The CPU workflow covers certificate checks, evaluation of saved Qwen/Olmo outputs, and count-panel statistical reconstruction. It recovers the Qwen/Olmo base-five point estimates; the original adjusted intervals in Tables~\ref{tab:primary}--\ref{tab:h2} retain their historical resampling schedule. Discovery, Phi, attention-validation, timed symbolic-search, and three-model prefix analyses are documented through aggregate tables and experimental protocols. The manifest marks original discovery and Phi trajectories, detailed attention logs, timed-search records, original resampling-index arrays, and three-model prefix record tables as \texttt{not\_in\_package}. Model generation and timed solver execution require the configurations and budgets specified in Appendices~\ref{app:implementation} and~\ref{app:solver}, respectively.

\subsection{Repository and versioning}\label{app:public-plan}
The companion repository records the artifact version, file digests, and the correspondence between paper results and replay outputs. Its manifest gives the evidence and reproduction scope of each archived component.

%% file: tables/readout.tex
\begin{tabular}{llrrrrrr}
\toprule
Model & Condition & Native & Final & Candidate & Parsed & CAP & EOS\\
\midrule
Qwen & Control & 67 & 67 & 67 & 75 & 43 & 77\\
 & Separation & 12 & 12 & 13 & 19 & 101 & 19\\
\addlinespace[2pt]
Olmo & Control & 3 & 52 & 57 & 57 & 63 & 57\\
 & Separation & 1 & 15 & 17 & 21 & 99 & 21\\
\addlinespace[2pt]
Phi$^{\dagger}$ & Control & 6 & 5 & 7 & 13 & 97 & 23\\
 & Separation & 3 & 2 & 3 & 8 & 100 & 20\\
\bottomrule
\end{tabular}

%% file: tables/exact.tex
\begin{tabular}{llrrr}
\toprule
Model & Target & Gain & Loss & $p_{\mathrm{Bonf},6}$\\
\midrule
Qwen & Visible six & 0 & 55 & 3.33e-16\\
Qwen & Common nine & 9 & 0 & 0.02344\\
Olmo & Visible six & 5 & 42 & 1.47e-07\\
Olmo & Common nine & 11 & 0 & 0.00586\\
Phi$^{\dagger}$ & Visible six & 2 & 5 & 1.00000\\
Phi$^{\dagger}$ & Common nine & 1 & 0 & 1.00000\\
\bottomrule
\end{tabular}

%% file: tables/program_classes.tex
\begin{tabular}{lllrrrr}
\toprule
Model & Audit & Condition & Success & Reference & Other 2 & Other length\\
\midrule
Qwen & Nine & Control & 0 & 0 & 0 & 0\\
Qwen & Nine & Separation & 9 & 0 & 9 & 0\\
Qwen & $H_2$ & Control & 0 & 0 & 0 & 0\\
Qwen & $H_2$ & Separation & 11 & 0 & 9 & 2\\
Olmo & Nine & Control & 0 & 0 & 0 & 0\\
Olmo & Nine & Separation & 11 & 0 & 10 & 1\\
Olmo & $H_2$ & Control & 0 & 0 & 0 & 0\\
Olmo & $H_2$ & Separation & 14 & 0 & 10 & 4\\
Phi$^{\dagger}$ & Nine & Control & 0 & 0 & 0 & 0\\
Phi$^{\dagger}$ & Nine & Separation & 1 & 0 & 1 & 0\\
Phi$^{\dagger}$ & $H_2$ & Control & 1 & 0 & 1 & 0\\
Phi$^{\dagger}$ & $H_2$ & Separation & 1 & 0 & 1 & 0\\
\bottomrule
\end{tabular}

%% file: tables/solution_groups.tex
\begin{tabular}{lllrrrrl}
\toprule
Model & $|\Sone(B_f)|$ & Target & $n$ & C & S & S $-$ C (pp) & Marginal 95\% interval \\
\midrule
Qwen & 1 & Visible six & 91 & 43 & 5 & -41.76 & [-51.65, -31.87] \\
Qwen & 1 & Common nine & 91 & 0 & 5 & +5.49 & [+1.10, +10.99] \\
Qwen & 1 & $H_2$ & 91 & 0 & 7 & +7.69 & [+3.30, +13.19] \\
Qwen & $>1$ & Visible six & 29 & 24 & 7 & -58.62 & [-75.86, -41.38] \\
Qwen & $>1$ & Common nine & 29 & 0 & 4 & +13.79 & [+3.45, +27.59] \\
Qwen & $>1$ & $H_2$ & 29 & 0 & 4 & +13.79 & [+3.45, +27.59] \\
Olmo & 1 & Visible six & 91 & 31 & 6 & -27.47 & [-37.36, -17.58] \\
Olmo & 1 & Common nine & 91 & 0 & 5 & +5.49 & [+1.10, +10.99] \\
Olmo & 1 & $H_2$ & 91 & 0 & 7 & +7.69 & [+2.20, +13.19] \\
Olmo & $>1$ & Visible six & 29 & 21 & 9 & -41.38 & [-65.52, -17.24] \\
Olmo & $>1$ & Common nine & 29 & 0 & 6 & +20.69 & [+6.90, +37.93] \\
Olmo & $>1$ & $H_2$ & 29 & 0 & 7 & +24.14 & [+10.34, +41.38] \\
Phi$^\dagger$ & 1 & Visible six & 91 & 2 & 2 & +0.00 & [-4.40, +4.40] \\
Phi$^\dagger$ & 1 & Common nine & 91 & 0 & 1 & +1.10 & [+0.00, +3.30] \\
Phi$^\dagger$ & 1 & $H_2$ & 91 & 1 & 1 & +0.00 & [-3.30, +3.30] \\
Phi$^\dagger$ & $>1$ & Visible six & 29 & 3 & 0 & -10.34 & [-24.14, +0.00] \\
Phi$^\dagger$ & $>1$ & Common nine & 29 & 0 & 0 & +0.00 & [+0.00, +0.00] \\
Phi$^\dagger$ & $>1$ & $H_2$ & 29 & 0 & 0 & +0.00 & [+0.00, +0.00] \\
\bottomrule
\end{tabular}

%% file: tables/token.tex
\begin{tabular}{lrrrr}
\toprule
Model & Mean & Median & Range & Zero difference\\
\midrule
Qwen & -0.0750 & 0.0 & [-3, 2] & 46/120\\
Olmo & -0.0750 & 0.0 & [-3, 2] & 46/120\\
Phi$^{\dagger}$ & -0.0083 & 0.0 & [-3, 3] & 55/120\\
\bottomrule
\end{tabular}

%% file: tables/token_strata.tex
\begin{tabular}{llrrr}
\toprule
Model & Token difference bin & $n$ & Visible $S-C$ (pp) & Marginal interval\\
\midrule
Qwen & \texttt{delta<=-2} & 5 & -80.00 & [-100.00, -40.00]\\
Qwen & \texttt{delta=-1} & 37 & -35.14 & [-51.35, -21.62]\\
Qwen & \texttt{delta=0} & 46 & -45.65 & [-60.87, -30.43]\\
Qwen & \texttt{delta=1} & 24 & -54.17 & [-75.00, -33.33]\\
Qwen & \texttt{delta>=2} & 8 & -50.00 & [-87.50, -12.50]\\
Olmo & \texttt{delta<=-2} & 5 & -40.00 & [-80.00, +0.00]\\
Olmo & \texttt{delta=-1} & 37 & -29.73 & [-45.95, -13.51]\\
Olmo & \texttt{delta=0} & 46 & -26.09 & [-43.48, -8.70]\\
Olmo & \texttt{delta=1} & 24 & -29.17 & [-50.00, -8.33]\\
Olmo & \texttt{delta>=2} & 8 & -62.50 & [-87.50, -25.00]\\
Phi$^{\dagger}$ & \texttt{delta<=-2} & 6 & +0.00 & [+0.00, +0.00]\\
Phi$^{\dagger}$ & \texttt{delta=-1} & 28 & -10.71 & [-25.00, +0.00]\\
Phi$^{\dagger}$ & \texttt{delta=0} & 55 & +0.00 & [-5.45, +5.45]\\
Phi$^{\dagger}$ & \texttt{delta=1} & 23 & +0.00 & [-13.04, +13.04]\\
Phi$^{\dagger}$ & \texttt{delta>=2} & 8 & +0.00 & [+0.00, +0.00]\\
\bottomrule
\end{tabular}

%% file: tables/extra.tex
\begin{tabular}{llrrrr}
\toprule
Model & Stratum & C (\%) & S (\%) & Difference (pp) & Marginal interval\\
\midrule
Qwen & Uniform & 48.10 & 11.98 & -36.12 & [-43.93, -28.36]\\
Qwen & Active & 0.60 & 2.76 & +2.16 & [-0.29, +5.34]\\
Olmo & Uniform & 34.30 & 12.58 & -21.72 & [-29.53, -13.70]\\
Olmo & Active & 0.42 & 2.01 & +1.59 & [-0.31, +4.22]\\
Phi$^{\dagger}$ & Uniform & 6.46 & 3.23 & -3.23 & [-7.84, +1.22]\\
Phi$^{\dagger}$ & Active & 0.60 & 0.00 & -0.60 & [-1.80, +0.00]\\
\bottomrule
\end{tabular}

%% file: tables/solver_anytime.tex
\begin{tabular}{rlrrr}
\toprule
CPU limit (s) & Condition & Visible six & Common nine & Unshown $H_2$\\
\midrule
0.1 & Control & 120/120 & 0/120 & 0/120\\
0.1 & Separation & 1/120 & 1/120 & 1/120\\
1 & Control & 120/120 & 0/120 & 0/120\\
1 & Separation & 32/120 & 28/120 & 29/120\\
10 & Control & 120/120 & 0/120 & 0/120\\
10 & Separation & 95/120 & 86/120 & 87/120\\
\bottomrule
\end{tabular}

%% file: tables/solver_candidates.tex
\begin{tabular}{rlrrr}
\toprule
Candidate limit & Condition & Visible six & Common nine & Unshown $H_2$\\
\midrule
1,000 & Control & 40/120 & 0/120 & 0/120\\
1,000 & Separation & 0/120 & 0/120 & 0/120\\
10,000 & Control & 120/120 & 0/120 & 0/120\\
10,000 & Separation & 0/120 & 0/120 & 0/120\\
100,000 & Control & 120/120 & 0/120 & 0/120\\
100,000 & Separation & 10/120 & 10/120 & 10/120\\
\bottomrule
\end{tabular}

%% file: tables/prefix.tex
\begin{tabular}{lrrrrrrr}
\toprule
Model & Tokens & C parsed & C final & C cand. & S parsed & S final & S cand.\\
\midrule
Qwen & 2,048 & 0 & 0 & 10 & 0 & 0 & 1\\
Qwen & 4,096 & 34 & 31 & 43 & 3 & 2 & 7\\
Qwen & 6,144 & 61 & 53 & 61 & 12 & 10 & 10\\
Qwen & 8,192 & 75 & 67 & 67 & 19 & 12 & 13\\
\addlinespace[2pt]
Olmo & 2,048 & 0 & 0 & 2 & 0 & 0 & 0\\
Olmo & 4,096 & 9 & 6 & 24 & 1 & 1 & 4\\
Olmo & 6,144 & 30 & 27 & 48 & 5 & 2 & 11\\
Olmo & 8,192 & 57 & 52 & 57 & 21 & 15 & 17\\
\addlinespace[2pt]
Phi$^{\dagger}$ & 2,048 & 0 & 0 & 0 & 0 & 0 & 0\\
Phi$^{\dagger}$ & 4,096 & 2 & 1 & 2 & 0 & 0 & 0\\
Phi$^{\dagger}$ & 6,144 & 7 & 4 & 6 & 2 & 1 & 2\\
Phi$^{\dagger}$ & 8,192 & 13 & 5 & 7 & 8 & 2 & 3\\
\bottomrule
\end{tabular}

%% file: r1_certificate_audit.tex
\FloatBarrier
\section{Additional Construction and Complete Panel Results}\label{app:r1archive}
\FloatBarrier
\subsection{Base variants and certificate persistence}\label{app:r1variants}
The extension retains all 120 original validation families. Common base counts three, four, five, and six give four, five, six, and seven displayed examples. The original base-five outputs are reused, and the other counts add 1,440 trajectories across two models and both conditions.

Reduced bases keep a fixed nonidentity anchor and follow an input-hash order, producing nested subsets. The extra base-six example satisfies both the one-rule control program and the family's two-rule reference. It is inserted immediately before the original changed example. The original changed-example positions one through six occur in 24/18/13/19/17/29 families. The new construction preserves the relative order of the original six examples.

The following table records certification for every variant. The same family has a one-rule solution in both conditions at base counts three and four. Its observations are reported below. The remaining families retain exact one-to-two certificates.
\begin{table}[ht]\centering\small
\caption{Certification of the paired specifications at each shared-base count. The four settings give four, five, six, and seven displayed examples.}
\begin{tabular}{rrrrrr}\toprule Common base&Displayed&One-to-two&One-to-one&Invalid&All families\\\midrule
3&4&119&1&0&120\\
4&5&119&1&0&120\\
5&6&120&0&0&120\\
6&7&120&0&0&120\\\bottomrule\end{tabular}
\end{table}

For base counts three, four, and five, 119 families have one nonidentity base example and one has two. At base six, 119 have one and one has three. The extra examples comprise 119 identity mappings and one rule-active mapping. Complete enumeration of the common bases gives $|\Sone(B)|=1,2,3$ in 90/13/17 families at base counts three and four, and 91/13/16 at five and six. Two families change their base solution count when examples are removed. After the control witness is added, only one family's one-rule solution count changes, from two at the reduced counts to one at counts five and six. The full control solution counts range from one to three, with median one, at every base count.

Within a family, the additional example excludes the original base, witnesses, and saved uniform and active inputs. Exclusion uses the pair consisting of family ID and input string. Sixteen added strings also occur among another family's audit inputs. Families use independent prompts and separate model calls. The standard-library verifier re-enumerates all 960 arm-specific certificates and checks the nested bases, reference execution, displayed order, and family-local exclusions.

\FloatBarrier
\subsection{A complete audit of the saved archive}\label{app:archiveexample}
The saved archive contains 1,008 tasks within the stated replacement language. We execute each saved reference and enumerate the complete one-rule solution set for its five displayed examples. The resulting certificates comprise 23 tasks with minimum length one and 226 with minimum length two. The remaining 759 have lower bound two and the upper bound supplied by their saved reference.

\begin{table}[ht]\centering\small
\caption{Certificate outcomes in the saved task archive. Reference length supplies an upper bound for tasks whose one-rule solution set is empty.}
\begin{tabular}{rrrrl}\toprule
Reference length & Tasks & Minimum 1 & Minimum 2 & Remaining bound\\\midrule
2 & 246 & 20 & 226 & ---\\
3 & 253 & 3 & 0 & 250 tasks: $2\leq\Lmin\leq3$\\
4 & 253 & 0 & 0 & 253 tasks: $2\leq\Lmin\leq4$\\
5 & 256 & 0 & 0 & 256 tasks: $2\leq\Lmin\leq5$\\\bottomrule
\end{tabular}\label{tab:archivecounts}\end{table}

A task selected by a fixed hash within the one-rule-shortcut stratum has reference
\[
\rep(\texttt{v},\texttt{zi});\quad
\rep(\texttt{z},\texttt{zv});\quad
\rep(\texttt{z},\texttt{b}).
\]
Its displayed examples also admit the single replacement $\rep(\texttt{v},\texttt{bvi})$. Both programs execute correctly on every row below. On input \texttt{z}, the three-rule reference returns \texttt{bv} and the one-rule program returns \texttt{z}.

\begin{table}[ht]\centering\small
\caption{A saved three-rule task with a certified one-rule solution. The examples specify behavior shared by the reference and its shorter alternative.}
\begin{tabular}{ll}\toprule Input & Output\\\midrule
\texttt{kybvd} & \texttt{kybbvid}\\
\texttt{uwah} & \texttt{uwah}\\
\texttt{gi} & \texttt{gi}\\
\texttt{ywh} & \texttt{ywh}\\
\texttt{vve} & \texttt{bvibvie}\\\bottomrule
\end{tabular}\label{tab:archiveexample}\end{table}

The archive audit and illustrative input are CPU executions of saved programs. All model inference in the count extension uses the 120 constructed validation families. The companion artifact includes the archive examples, reference programs, short witnesses, and certificate results for direct replay.

\FloatBarrier
\subsection{Evaluation targets and statistical reconstruction}\label{app:r1completedpanel}
The count panel contains 1,920 stored outputs: 1,440 new outputs and 480 original base-five outputs. It covers every combination of two models, 120 families, four base counts, and two conditions. The certified one-to-two sets contain 119/119/120/120 families at base counts three/four/five/six; the full evaluated sets each contain 120. The fixed common-certificate set for cross-count comparisons contains 119 families.

The original common-nine target consists of the five original base examples and all four witnesses. It stays fixed across base counts. The wholly unshown pair contains the two witnesses absent from both prompts. A further recorded target contains the original three examples absent from the arm's base-five prompt. These target sets are stored explicitly. Uniform and ACTIVE each contain 32 reference-generated inputs per family, and their scores are within-family hit fractions.

The original reference satisfies every displayed specification in the 480 paired variants. Its outputs agree with all 7,680 saved broader labels, as do those of the stored two-rule separation witnesses. All 3,840 ACTIVE inputs sequentially activate both reference rules. Executing the 1,920 selected programs on these inputs gives 122,880 program-input evaluations.

The selector yields an unparsable final object for 1,207 of the 1,920 outputs, including 899 new outputs. These records have zero success under the fixed scoring rule. Every requested output is present. The new trajectories end at EOS in 544 cases and at the token cap in 896 cases. The companion artifact retains the generated text, selected-program status, and family-level outcomes for all records.

All extension intervals below use 20,000 paired-family bootstrap resamples with PCG64 seed 20261107. Individual contrasts use pointwise 95\% intervals. The six planned cross-count displayed-success contrasts use percentile quantiles $1/240$ and $239/240$. Base-five point estimates match the original validation. The original primary and $H_2$ intervals retain their original 50,000-resample adjustment in Tables~\ref{tab:primary} and~\ref{tab:h2}; the tables below present the count panel under its common statistical schedule.

\FloatBarrier
\subsection{Displayed and wholly unshown outcomes}\label{app:r1outcomes}
The next two tables show the count-specific certified comparisons and the corresponding full evaluated cohorts. The certified comparisons have negative displayed-task changes and positive unshown-pair changes throughout.
\begin{table}[ht]\centering\small
\caption{Qwen3-8B across example counts. C/S gives displayed-task successes. The displayed and unshown-pair columns summarize certified one-to-two pairs; the final column summarizes the full evaluated cohort.}
\begingroup\setlength{\tabcolsep}{3pt}\begin{tabular}{rrrrrr}\toprule Base&Primary N&C/S&Displayed $\Delta$ [CI]&Unshown pair $\Delta$ [CI]&All-legal displayed $\Delta$ [CI]\\\midrule
3&119&59/15&$-36.97\, [-46.22, -27.73]$&$+13.45\, [+7.56, +20.17]$&$-36.67\, [-45.83, -27.50]$\\
4&119&61/15&$-38.66\, [-47.90, -28.57]$&$+10.92\, [+5.88, +16.81]$&$-39.17\, [-48.33, -29.17]$\\
5&120&67/12&$-45.83\, [-55.00, -36.67]$&$+9.17\, [+4.17, +15.00]$&$-45.83\, [-55.00, -36.67]$\\
6&120&68/19&$-40.83\, [-50.00, -31.67]$&$+12.50\, [+6.67, +18.33]$&$-40.83\, [-50.00, -31.67]$\\
\bottomrule\end{tabular}\endgroup\label{tab:r1completeprimary}\end{table}
\begin{table}[ht]\centering\small
\caption{Olmo-3-7B-Think across example counts. Displayed completion decreases and unshown-pair success increases in each certified comparison.}
\begingroup\setlength{\tabcolsep}{3pt}\begin{tabular}{rrrrrr}\toprule Base&Primary N&C/S&Displayed $\Delta$ [CI]&Unshown pair $\Delta$ [CI]&All-legal displayed $\Delta$ [CI]\\\midrule
3&119&53/14&$-32.77\, [-42.86, -22.69]$&$+8.40\, [+3.36, +13.45]$&$-33.33\, [-43.33, -23.33]$\\
4&119&47/19&$-23.53\, [-34.45, -12.61]$&$+11.76\, [+5.88, +17.65]$&$-23.33\, [-34.17, -12.50]$\\
5&120&52/15&$-30.83\, [-40.83, -20.83]$&$+11.67\, [+6.67, +17.50]$&$-30.83\, [-40.83, -20.83]$\\
6&120&54/13&$-34.17\, [-43.33, -25.00]$&$+9.17\, [+4.17, +15.00]$&$-34.17\, [-43.33, -25.00]$\\
\bottomrule\end{tabular}\endgroup\label{tab:r1completeprimaryolmo}\end{table}
\FloatBarrier
\subsection{Paired changes relative to base five}\label{app:r1crosscount}
For each family in the 119-family common-certificate set, we compute the displayed-success contrast at a new base count and subtract the contrast for the same family's base-five outputs. Resampling preserves these within-family comparisons. The six adjusted intervals span zero.
\begin{table}[ht]\centering\small

\caption{Paired changes in the displayed-task contrast relative to base five. The intervals use the adjustment across the six planned comparisons.}
\begin{tabular}{lrr}\toprule Model&New base count&$\Delta_b-\Delta_5$ [adjusted CI]\\\midrule

Qwen3-8B&3&$+8.40\, [-7.56, +23.53]$\\

Qwen3-8B&4&$+6.72\, [-8.13, +21.85]$\\

Qwen3-8B&6&$+4.20\, [-8.40, +16.81]$\\

Olmo-3-7B-Think&3&$-1.68\, [-17.37, +14.29]$\\

Olmo-3-7B-Think&4&$+7.56\, [-10.08, +26.05]$\\

Olmo-3-7B-Think&6&$-3.36\, [-19.33, +13.45]$\\

\bottomrule\end{tabular}\end{table}
\FloatBarrier
\subsection{Complete endpoint panel}\label{app:r1endpoints}
Displayed success tests the arm's own specification. One-rule delivery records a selected one-rule solution to that specification. Correct-candidate availability records whether a valid task solution occurs in the saved output. The legacy parser applies the original native extraction rule. Common-nine, the wholly unshown pair, the original three unshown examples, and the two broader strata use the fixed targets defined above. The following tables report each endpoint for both models, with values rounded to one decimal place; the supplied CSVs retain full precision.
\begin{table}[ht]\centering\footnotesize
\caption{Qwen3-8B endpoint changes across example counts. Values give separation-minus-control changes in percentage points and pointwise 95\% paired-family intervals.}
\begingroup\setlength{\tabcolsep}{3pt}\begin{tabular}{lrrrr}\toprule Endpoint&Base3&Base4&Base5&Base6\\\midrule
Displayed specification&-37.0 [-46.2,-27.7]&-38.7 [-47.9,-28.6]&-45.8 [-55.0,-36.7]&-40.8 [-50.0,-31.7]\\
One-rule delivered&-26.9 [-35.3,-19.3]&-29.4 [-37.8,-21.8]&-37.5 [-46.7,-29.2]&-44.2 [-53.3,-35.0]\\
Common-nine&+10.1 [5.0,16.0]&+8.4 [4.2,13.4]&+7.5 [3.3,12.5]&+10.8 [5.8,16.7]\\
Unshown pair&+13.4 [7.6,20.2]&+10.9 [5.9,16.8]&+9.2 [4.2,15.0]&+12.5 [6.7,18.3]\\
Original three unshown&+12.6 [6.7,18.5]&+10.1 [5.0,16.0]&+8.3 [4.2,13.3]&+10.8 [5.8,16.7]\\
UNIFORM&-23.2 [-30.8,-15.5]&-27.3 [-34.6,-20.0]&-36.1 [-43.9,-28.3]&-31.0 [-38.9,-23.0]\\
ACTIVE&+5.3 [1.7,9.5]&+3.4 [0.1,7.2]&+2.2 [-0.3,5.3]&+4.5 [1.2,8.6]\\
Raw valid candidate&-37.0 [-46.2,-27.7]&-38.7 [-47.9,-29.4]&-45.0 [-54.2,-35.8]&-38.3 [-47.5,-29.2]\\
Legacy parser&-37.0 [-46.2,-27.7]&-38.7 [-47.9,-28.6]&-45.8 [-55.0,-36.7]&-40.8 [-50.0,-31.7]\\
\bottomrule\end{tabular}\endgroup\end{table}
\begin{table}[ht]\centering\footnotesize
\caption{Olmo-3-7B-Think endpoint changes across example counts. Values give separation-minus-control changes in percentage points and pointwise 95\% paired-family intervals.}
\begingroup\setlength{\tabcolsep}{3pt}\begin{tabular}{lrrrr}\toprule Endpoint&Base3&Base4&Base5&Base6\\\midrule
Displayed specification&-32.8 [-42.9,-22.7]&-23.5 [-34.5,-12.6]&-30.8 [-40.8,-20.8]&-34.2 [-43.3,-25.0]\\
One-rule delivered&-21.0 [-28.6,-14.3]&-24.4 [-31.9,-16.8]&-27.5 [-35.8,-20.0]&-33.3 [-41.7,-25.0]\\
Common-nine&+6.7 [2.5,11.8]&+9.2 [4.2,14.3]&+9.2 [4.2,14.2]&+8.3 [3.3,13.3]\\
Unshown pair&+8.4 [3.4,13.4]&+11.8 [5.9,17.6]&+11.7 [6.7,17.5]&+9.2 [4.2,15.0]\\
Original three unshown&+8.4 [3.4,13.4]&+11.8 [5.9,17.6]&+10.0 [5.0,15.8]&+8.3 [3.3,13.3]\\
UNIFORM&-23.5 [-31.9,-15.0]&-18.5 [-27.3,-9.7]&-21.7 [-29.6,-13.8]&-27.9 [-35.5,-20.4]\\
ACTIVE&+4.7 [1.2,8.7]&+4.5 [1.3,8.2]&+1.6 [-0.3,4.2]&+1.5 [-0.5,4.2]\\
Raw valid candidate&-32.8 [-42.9,-22.7]&-27.7 [-38.7,-16.8]&-33.3 [-43.3,-23.3]&-37.5 [-46.7,-28.3]\\
Legacy parser&-0.8 [-2.5,0.0]&-0.8 [-2.5,0.0]&-1.7 [-5.0,1.7]&-2.5 [-5.8,0.0]\\
\bottomrule\end{tabular}\endgroup\end{table}
\FloatBarrier
\subsection{The one-to-one variants}\label{app:r1oneone}
One family retains a one-rule solution under both conditions at each of the reduced base counts. The table gives the four observed model-by-count pairs for this family. The programs and examples are included in the same replay files as the certified one-to-two cases. The reference applies \texttt{f}$\to$\texttt{fdd}, followed by \texttt{db}$\to$\texttt{c}. The reduced separation specification admits \texttt{b}$\to$\texttt{dc}. Restoring \texttt{iyxbdc}$\to$\texttt{iyxbdc} removes that solution, which predicts \texttt{iyxdcdc} on this input. The file \texttt{worked\_case.json} contains the complete specifications and the replay entry verifies each certificate.
\begin{table}[ht]\centering\small\caption{Observed outcomes for the family with one-to-one variants. Entries give the Control/Separation success indicators for each model and base count.}
\begin{tabular}{lrrr}\toprule Model&Base&Displayed C/S&Unshown pair C/S\\\midrule

Qwen3-8B&3&1/1&0/1\\

Qwen3-8B&4&1/0&1/0\\

Olmo-3-7B-Think&3&1/0&1/1\\

Olmo-3-7B-Think&4&1/1&1/1\\

\bottomrule\end{tabular}\end{table}
\FloatBarrier
\subsection{Summary of the completed observations}
Sections~\ref{sec:primary}--\ref{sec:broader} summarize the observed task-completion, witness, and broader-input patterns. The tables above provide all certified comparisons, full-cohort results, and one-to-one observations; Appendix~\ref{app:evidence} identifies their executable records.

%% file: main.bbl
\begin{thebibliography}{31}
\providecommand{\natexlab}[1]{#1}
\providecommand{\url}[1]{\texttt{#1}}
\expandafter\ifx\csname urlstyle\endcsname\relax
  \providecommand{\doi}[1]{doi: #1}\else
  \providecommand{\doi}{doi: \begingroup \urlstyle{rm}\Url}\fi

\bibitem[Alur et~al.(2013)Alur, Bod{\'\i}k, Juniwal, Martin, Raghothaman,
  Seshia, Singh, Solar-Lezama, Torlak, and Udupa]{alur2013}
Rajeev Alur, Rastislav Bod{\'\i}k, Garvit Juniwal, Milo M.~K. Martin, Mukund
  Raghothaman, Sanjit~A. Seshia, Rishabh Singh, Armando Solar-Lezama, Emina
  Torlak, and Abhishek Udupa.
\newblock Syntax-guided synthesis.
\newblock In \emph{Formal Methods in Computer-Aided Design}, pages 1--17, 2013.
\newblock \doi{10.1109/FMCAD.2013.6679385}.

\bibitem[Angluin(1987)]{angluin1987}
Dana Angluin.
\newblock Learning regular sets from queries and counterexamples.
\newblock \emph{Information and Computation}, 75\penalty0 (2):\penalty0
  87--106, 1987.
\newblock \doi{10.1016/0890-5401(87)90052-6}.

\bibitem[Balog et~al.(2017)Balog, Gaunt, Brockschmidt, Nowozin, and
  Tarlow]{balog2017}
Matej Balog, Alexander~L. Gaunt, Marc Brockschmidt, Sebastian Nowozin, and
  Daniel Tarlow.
\newblock {DeepCoder}: Learning to write programs, 2017.
\newblock URL \url{https://openreview.net/forum?id=rkE3y85ee}.

\bibitem[Batzoglou(2026)]{replayscm}
Serafim Batzoglou.
\newblock {ReplaySCM}: A benchmark for executable causal mechanism induction
  from interventions.
\newblock arXiv:2605.08197v1, 2026.
\newblock URL \url{https://arxiv.org/abs/2605.08197v1}.

\bibitem[Bowman and Dahl(2021)]{bowman2021}
Samuel~R. Bowman and George Dahl.
\newblock What will it take to fix benchmarking in natural language
  understanding?
\newblock In \emph{Proceedings of the 2021 Conference of the North American
  Chapter of the Association for Computational Linguistics: Human Language
  Technologies}, pages 4843--4855, 2021.
\newblock \doi{10.18653/v1/2021.naacl-main.385}.

\bibitem[Caldarella et~al.(2026)Caldarella, Talon, Aljundi, Ricci, and
  Mancini]{thinkingpast}
Simone Caldarella, Davide Talon, Rahaf Aljundi, Elisa Ricci, and Massimiliano
  Mancini.
\newblock Thinking past the answer: Evaluating harmful overthinking in large
  reasoning models.
\newblock arXiv:2606.02835v1, 2026.
\newblock URL \url{https://arxiv.org/abs/2606.02835v1}.

\bibitem[Devlin et~al.(2017)Devlin, Uesato, Bhupatiraju, Singh, rahman Mohamed,
  and Kohli]{devlin2017}
Jacob Devlin, Jonathan Uesato, Surya Bhupatiraju, Rishabh Singh, Abdel rahman
  Mohamed, and Pushmeet Kohli.
\newblock {RobustFill}: Neural program learning under noisy {I/O}.
\newblock In \emph{Proceedings of the 34th International Conference on Machine
  Learning}, volume~70 of \emph{Proceedings of Machine Learning Research},
  pages 990--998, 2017.
\newblock URL \url{https://proceedings.mlr.press/v70/devlin17a.html}.

\bibitem[Dror et~al.(2018)Dror, Baumer, Shlomov, and Reichart]{dror2018}
Rotem Dror, Gili Baumer, Segev Shlomov, and Roi Reichart.
\newblock The hitchhiker's guide to testing statistical significance in natural
  language processing.
\newblock In \emph{Proceedings of the 56th Annual Meeting of the Association
  for Computational Linguistics (Volume 1: Long Papers)}, pages 1383--1392,
  2018.
\newblock \doi{10.18653/v1/P18-1128}.

\bibitem[Efron(1979)]{efron1979}
Bradley Efron.
\newblock Bootstrap methods: Another look at the jackknife.
\newblock \emph{The Annals of Statistics}, 7\penalty0 (1):\penalty0 1--26,
  1979.
\newblock \doi{10.1214/aos/1176344552}.

\bibitem[Gardner et~al.(2020)Gardner, Artzi, Basmov, Berant, Bogin, Chen,
  Dasigi, Dua, Elazar, Gottumukkala, Gupta, Hajishirzi, Ilharco, Khashabi, Lin,
  Liu, Liu, Mulcaire, Ning, Singh, Smith, Subramanian, Tsarfaty, Wallace,
  Zhang, and Zhou]{gardner2020}
Matt Gardner, Yoav Artzi, Victoria Basmov, Jonathan Berant, Ben Bogin, Sihao
  Chen, Pradeep Dasigi, Dheeru Dua, Yanai Elazar, Ananth Gottumukkala, Nitish
  Gupta, Hannaneh Hajishirzi, Gabriel Ilharco, Daniel Khashabi, Kevin Lin,
  Jiangming Liu, Nelson~F. Liu, Phoebe Mulcaire, Qiang Ning, Sameer Singh,
  Noah~A. Smith, Sanjay Subramanian, Reut Tsarfaty, Eric Wallace, Ally Zhang,
  and Ben Zhou.
\newblock Evaluating models' local decision boundaries via contrast sets.
\newblock In \emph{Findings of the Association for Computational Linguistics:
  EMNLP 2020}, pages 1307--1323, 2020.
\newblock \doi{10.18653/v1/2020.findings-emnlp.117}.

\bibitem[Geirhos et~al.(2020)Geirhos, Jacobsen, Michaelis, Zemel, Brendel,
  Bethge, and Wichmann]{geirhos2020}
Robert Geirhos, J{\"o}rn-Henrik Jacobsen, Claudio Michaelis, Richard Zemel,
  Wieland Brendel, Matthias Bethge, and Felix~A. Wichmann.
\newblock Shortcut learning in deep neural networks.
\newblock \emph{Nature Machine Intelligence}, 2:\penalty0 665--673, 2020.
\newblock \doi{10.1038/s42256-020-00257-z}.

\bibitem[Gulwani(2011)]{gulwani2011}
Sumit Gulwani.
\newblock Automating string processing in spreadsheets using input-output
  examples.
\newblock In \emph{Proceedings of the 38th ACM SIGPLAN-SIGACT Symposium on
  Principles of Programming Languages}, pages 317--330, 2011.
\newblock \doi{10.1145/1926385.1926423}.

\bibitem[Gulwani et~al.(2017)Gulwani, Polozov, and Singh]{gulwani2017}
Sumit Gulwani, Oleksandr Polozov, and Rishabh Singh.
\newblock Program synthesis.
\newblock \emph{Foundations and Trends in Programming Languages}, 4\penalty0
  (1--2):\penalty0 1--119, 2017.
\newblock \doi{10.1561/2500000010}.

\bibitem[Gururangan et~al.(2018)Gururangan, Swayamdipta, Levy, Schwartz,
  Bowman, and Smith]{gururangan2018}
Suchin Gururangan, Swabha Swayamdipta, Omer Levy, Roy Schwartz, Samuel~R.
  Bowman, and Noah~A. Smith.
\newblock Annotation artifacts in natural language inference data.
\newblock In \emph{Proceedings of NAACL-HLT, Volume 2 (Short Papers)}, pages
  107--112, 2018.
\newblock \doi{10.18653/v1/N18-2017}.

\bibitem[Hong et~al.(2021)Hong, Dohan, Singh, Sutton, and
  Zaheer]{hong2021latent}
Joey Hong, David Dohan, Rishabh Singh, Charles Sutton, and Manzil Zaheer.
\newblock Latent programmer: Discrete latent codes for program synthesis.
\newblock In \emph{Proceedings of the 38th International Conference on Machine
  Learning}, volume 139 of \emph{Proceedings of Machine Learning Research},
  pages 4308--4318. PMLR, 2021.
\newblock URL \url{https://proceedings.mlr.press/v139/hong21a.html}.

\bibitem[Jacobs and Wallach(2021)]{jacobs2021}
Abigail~Z. Jacobs and Hanna Wallach.
\newblock Measurement and fairness.
\newblock In \emph{Proceedings of the 2021 ACM Conference on Fairness,
  Accountability, and Transparency}, pages 375--385, 2021.
\newblock \doi{10.1145/3442188.3445901}.

\bibitem[Kim et~al.(2021)Kim, Hu, D'Antoni, and Reps]{kim2021semgus}
Jinwoo Kim, Qinheping Hu, Loris D'Antoni, and Thomas Reps.
\newblock Semantics-guided synthesis.
\newblock \emph{Proceedings of the ACM on Programming Languages}, 5\penalty0
  (POPL):\penalty0 1--32, 2021.
\newblock \doi{10.1145/3434311}.

\bibitem[McCoy et~al.(2019)McCoy, Pavlick, and Linzen]{mccoy2019}
R.~Thomas McCoy, Ellie Pavlick, and Tal Linzen.
\newblock Right for the wrong reasons: Diagnosing syntactic heuristics in
  natural language inference.
\newblock In \emph{Proceedings of the 57th Annual Meeting of the Association
  for Computational Linguistics}, pages 3428--3448, 2019.
\newblock \doi{10.18653/v1/P19-1334}.

\bibitem[McNemar(1947)]{mcnemar1947}
Quinn McNemar.
\newblock Note on the sampling error of the difference between correlated
  proportions or percentages.
\newblock \emph{Psychometrika}, 12\penalty0 (2):\penalty0 153--157, 1947.
\newblock \doi{10.1007/BF02295996}.

\bibitem[Naik et~al.(2026{\natexlab{a}})Naik, Agrawal, Prakam, Mathur,
  Kapadnis, An, Marr, Rose, and Mortensen]{pbebench}
Atharva Naik, Darsh Agrawal, Prakam, Yash Mathur, Manav Kapadnis, Yuwei An,
  Clayton Marr, Carolyn Rose, and David Mortensen.
\newblock {PBEBench}: A multi-step programming by examples reasoning benchmark
  inspired by historical linguistics.
\newblock In \emph{Findings of the Association for Computational Linguistics:
  ACL 2026}, pages 8877--8918, 2026{\natexlab{a}}.
\newblock \doi{10.18653/v1/2026.findings-acl.432}.

\bibitem[Naik et~al.(2026{\natexlab{b}})Naik, Mathur, Prakam, Rose, and
  Mortensen]{reacomp}
Atharva Naik, Yash Mathur, Prakam, Carolyn Rose, and David Mortensen.
\newblock {ReaComp}: Compiling {LLM} reasoning into symbolic solvers for
  efficient program synthesis.
\newblock arXiv:2605.05485v1, 2026{\natexlab{b}}.
\newblock URL \url{https://arxiv.org/abs/2605.05485v1}.

\bibitem[Ozelim and Zenil(2026)]{ficl}
Luan Ozelim and Hector Zenil.
\newblock Measuring in-context algorithmic reasoning in language models against
  an exact {Bayes}-optimal reference.
\newblock arXiv:2608.01575v2, 2026.
\newblock URL \url{https://arxiv.org/html/2608.01575v2}.

\bibitem[Ribeiro et~al.(2020)Ribeiro, Wu, Guestrin, and Singh]{ribeiro2020}
Marco~Tulio Ribeiro, Tongshuang Wu, Carlos Guestrin, and Sameer Singh.
\newblock Beyond accuracy: Behavioral testing of {NLP} models with {CheckList}.
\newblock In \emph{Proceedings of the 58th Annual Meeting of the Association
  for Computational Linguistics}, pages 4902--4912, 2020.
\newblock \doi{10.18653/v1/2020.acl-main.442}.

\bibitem[Shi et~al.(2024)Shi, Hong, Deng, Yin, Zaheer, and
  Sutton]{shi2024exedec}
Kensen Shi, Joey Hong, Yinlin Deng, Pengcheng Yin, Manzil Zaheer, and Charles
  Sutton.
\newblock Exedec: Execution decomposition for compositional generalization in
  neural program synthesis, 2024.
\newblock URL \url{https://arxiv.org/abs/2307.13883v2}.

\bibitem[Solar-Lezama(2008)]{solar2008}
Armando Solar-Lezama.
\newblock \emph{Program Synthesis by Sketching}.
\newblock PhD thesis, University of California, Berkeley, 2008.
\newblock URL
  \url{https://www2.eecs.berkeley.edu/Pubs/TechRpts/2008/EECS-2008-176.html}.

\bibitem[Solar-Lezama et~al.(2006)Solar-Lezama, Tancau, Bod{\'\i}k, Seshia, and
  Saraswat]{solar2006}
Armando Solar-Lezama, Liviu Tancau, Rastislav Bod{\'\i}k, Sanjit~A. Seshia, and
  Vijay~A. Saraswat.
\newblock Combinatorial sketching for finite programs.
\newblock In \emph{Proceedings of the 12th International Conference on
  Architectural Support for Programming Languages and Operating Systems}, pages
  404--415, 2006.
\newblock URL
  \url{https://people.eecs.berkeley.edu/~sseshia/pubs/b2hd-solar-asplos06.html}.

\bibitem[Srivastava et~al.(2023)Srivastava, Rastogi, Rao,
  et~al.]{srivastava2023}
Aarohi Srivastava, Abhinav Rastogi, Abhishek Rao, et~al.
\newblock Beyond the imitation game: Quantifying and extrapolating the
  capabilities of language models.
\newblock \emph{Transactions on Machine Learning Research}, 2023.
\newblock URL \url{https://openreview.net/forum?id=uyTL5Bvosj}.

\bibitem[{Team Olmo} et~al.(2025){Team Olmo}, Ettinger, Bertsch, Kuehl, Graham,
  Heineman, Groeneveld, et~al.]{olmo2025}
{Team Olmo}, Allyson Ettinger, Amanda Bertsch, Bailey Kuehl, David Graham,
  David Heineman, Dirk Groeneveld, et~al.
\newblock {Olmo 3}.
\newblock arXiv:2512.13961, 2025.
\newblock URL \url{https://arxiv.org/abs/2512.13961}.

\bibitem[Wei et~al.(2022)Wei, Wang, Schuurmans, Bosma, Ichter, Xia, Chi, Le,
  and Zhou]{wei2022}
Jason Wei, Xuezhi Wang, Dale Schuurmans, Maarten Bosma, Brian Ichter, Fei Xia,
  Ed~H. Chi, Quoc~V. Le, and Denny Zhou.
\newblock Chain-of-thought prompting elicits reasoning in large language
  models.
\newblock In \emph{Advances in Neural Information Processing Systems},
  volume~35, pages 24824--24837, 2022.
\newblock URL
  \url{https://proceedings.neurips.cc/paper/2022/hash/9d5609613524ecf4f15af0f7b31abca4-Abstract-Conference.html}.

\bibitem[Xu et~al.(2025)Xu, Peng, Awadalla, Chen, Chen, Gao, Kim, Li, Ren,
  Shen, Wang, Xu, Gao, and Chen]{phi2025}
Haoran Xu, Baolin Peng, Hany Awadalla, Dongdong Chen, Yen-Chun Chen, Mei Gao,
  Young~Jin Kim, Yunsheng Li, Liliang Ren, Yelong Shen, Shuohang Wang, Weijian
  Xu, Jianfeng Gao, and Weizhu Chen.
\newblock {Phi-4-Mini-Reasoning}: Exploring the limits of small reasoning
  language models in math.
\newblock arXiv:2504.21233, 2025.
\newblock URL \url{https://arxiv.org/abs/2504.21233}.

\bibitem[Yang et~al.(2025)Yang, Li, Yang, Zhang, Hui, et~al.]{qwen2025}
An~Yang, Anfeng Li, Baosong Yang, Beichen Zhang, Binyuan Hui, et~al.
\newblock {Qwen3} technical report.
\newblock arXiv:2505.09388, 2025.
\newblock URL \url{https://arxiv.org/abs/2505.09388}.

\end{thebibliography}
